\documentclass{aa}

\usepackage[utf8]{inputenc}
\usepackage[T1]{fontenc}

\usepackage[varg]{txfonts}

\usepackage{graphicx}	
\usepackage{amsmath}	
\usepackage{amssymb}	
\usepackage{threeparttable}
\usepackage{booktabs}
\usepackage{verbatim}
\usepackage{caption}
\usepackage[normalem]{ulem}
\usepackage[usenames, dvipsnames]{color}

\definecolor{linkcolor}{rgb}{0.6,0,0} 
\definecolor{citecolor}{rgb}{0,0,0.75}
\definecolor{urlcolor}{rgb}{0.12,0.46,0.7}
\usepackage[breaklinks, colorlinks, urlcolor=urlcolor, linkcolor=linkcolor, citecolor=citecolor]{hyperref}
\hypersetup{linktocpage}
\usepackage[all]{hypcap}

\begin{document}

\title{KiDS-1000: Combined halo-model cosmology constraints from galaxy abundance, galaxy clustering and galaxy-galaxy lensing}

\author{Andrej Dvornik\inst{1}
\and Catherine Heymans\inst{2,1}
\and Marika Asgari\inst{3,2}
\and Constance Mahony\inst{1}
\and Benjamin Joachimi\inst{4}
\and Maciej Bilicki\inst{5}
\and Elisa Chisari\inst{6}
\and Hendrik Hildebrandt\inst{1}
\and Henk Hoekstra\inst{7}
\and Harry Johnston\inst{6}
\and Konrad Kuijken\inst{7}
\and Alexander Mead\inst{8,1}
\and Hironao Miyatake\inst{9,10,11}
\and Takahiro Nishimichi\inst{12,10}
\and Robert Reischke\inst{1}
\and Sandra Unruh\inst{13,1}
\and Angus H. Wright\inst{1}
}
\institute{Ruhr University Bochum, Faculty of Physics and Astronomy, Astronomical Institute (AIRUB), German Centre for Cosmological Lensing, 44780 Bochum, Germany\\
\email{dvornik@astro.rub.de}
\and Institute for Astronomy, University of Edinburgh, Royal Observatory, Blackford Hill, Edinburgh EH9 3HJ, UK
\and  E. A. Milne Centre, University of Hull, Cottingham Road, Hull, HU6 7RX, UK
\and Department of Physics and Astronomy, University College London, Gower Street, London WC1E 6BT, UK
\and Center for Theoretical Physics, Polish Academy of Sciences, al. Lotników 32/46, 02-668 Warsaw, Poland
\and Institute for Theoretical Physics, Utrecht University, 3584CC Utrecht, The Netherlands
\and Leiden Observatory, Leiden University, P.O.Box 9513, 2300RA Leiden, The Netherlands
\and Department of Computer Science, University of British Columbia, 6224 Agricultural Road, Vancouver, BC V6T 1Z1, Canada 
\and Kobayashi-Maskawa Institute for the Origin of Particles and the Universe (KMI), Nagoya University, Nagoya 464-8602, Japan
\and Institute for Advanced Research, Nagoya University, Nagoya 464-8601, Japan
\and Kavli Institute for the Physics and Mathematics of the Universe (WPI), The University of Tokyo Institutes for Advanced Study, The University of Tokyo, Chiba 277-8583, Japan
\and Center for Gravitational Physics and Quantum Information, Yukawa Institute for Theoretical Physics, Kyoto University, Kyoto 606-8502, Japan
\and Argelander-Institut für Astronomie, Universität Bonn, Auf dem Hügel 71, D-53121 Bonn, Germany}

\date{Received ???; Accepted ???}

\keywords{gravitational lensing: weak -- methods: statistical -- cosmological parameters -- galaxies: haloes -- dark matter -- large-scale structure of Universe.}

\titlerunning{KiDS-1000: 2$\times$2 point cosmology with SMF}
\authorrunning{A. Dvornik et al.}

\abstract{

We present constraints on the flat $\Lambda$CDM cosmological model through a joint analysis of galaxy abundance, galaxy clustering and galaxy-galaxy lensing observables with the Kilo-Degree Survey. Our theoretical model combines a flexible conditional stellar mass function, to describe the galaxy-halo connection, with a cosmological N-body simulation-calibrated halo model to describe the non-linear matter field. Our magnitude-limited bright galaxy sample combines 9-band optical-to-near-infrared photometry with an extensive and complete spectroscopic training sample to provide accurate redshift and stellar mass estimates.  Our faint galaxy sample provides a background of accurately calibrated lensing measurements.  We constrain the structure growth parameter $S_8 = \sigma_8 \sqrt{\Omega_{\mathrm{m}}/0.3} = 0.773^{+0.028}_{-0.030}$, and the matter density parameter $\Omega_{\mathrm{m}} = 0.290^{+0.021}_{-0.017}$. The galaxy-halo connection model adopted in the work is shown to be in agreement with previous studies. Our constraints on cosmological parameters are comparable to, and consistent with, joint `$3\times2{\mathrm{pt}}$’ clustering-lensing analyses that additionally include a cosmic shear observable.  This analysis therefore brings attention to the significant constraining power in the often-excluded non-linear scales for galaxy clustering and galaxy-galaxy lensing observables. By adopting a theoretical model that accounts for non-linear halo bias, halo exclusion, scale-dependent galaxy bias and the impact of baryon feedback, this work demonstrates the potential and a way forward to include non-linear scales in cosmological analyses. Varying the width of the satellite galaxy distribution with an additional parameter yields a strong preference for sub-Poissonian variance, improving the goodness of fit by 0.18 in reduced $\chi^{2}$ value (and increasing the $p$-value by 0.25) compared to a fixed Poisson distribution.}

\maketitle

\section{Introduction}
\label{sec:intro}

In the last quarter of the century we have seen the rise and establishment of the concordance cosmological model which describes the formation and subsequent evolution of the cosmic structure. In this concordance model the Universe at present time is modelled as a flat, cold dark matter (CDM) and cosmological constant ($\Lambda$) dominated Universe with a negligible contribution from neutrinos, and gravity described by the law of General Relativity. Furthermore, the initial power spectrum of density fluctuations is assumed to be a single power-law. Such $\Lambda$CDM models are described using only six free parameters, which govern the energy densities of baryons, $\Omega_{\mathrm{b}}$, and cold dark matter, $\Omega_{\mathrm{dm}}$, the spectral index, $n_\mathrm{s}$, and normalisation, $\sigma_8$, of the density perturbations power spectrum, the Hubble parameter,  $H_0$, and the optical depth of reionisation. The flat geometry, implying that $\Omega_{\Lambda} = 1 - \Omega_{\mathrm{b}} - \Omega_{\mathrm{dm}}$, is strongly supported by high precision early-Universe measurements of the cosmic microwave background (CMB) temperature fluctuations combined with supernova and/or baryon acoustic oscillation distance measurements \citep{PlanckCollaboration2018,Scolnic2018,eboss2021}. Formally the models also include the total neutrino mass, but the value of the parameter is too small for the current precision of observations \citep{Gerbino2018}.

A powerful probe of late-time cosmology is the large-scale distribution of galaxies. Even though the stars contribute negligibly to the overall energy density of the Universe, the light from stars in galaxies can be used to trace the evolution of the underlying distribution of dark matter in two complementary ways. Firstly, the light-path from distant galaxies is impacted by the distribution of foreground mass. This `gravitational lensing' effect leads to a correlation between the observed shapes of galaxies, commonly referred to as \textit{cosmic shear}.  This observable can be used to probe the statistical properties of the total matter distribution in the Universe, typically quantified through the shape and amplitude of the matter power spectrum \citep{Heymans2013, Hikage2018, Hamana2019, Asgari2020, Secco2022, Amon2021, Busch2022}. Secondly, galaxies are expected to reside within dark matter haloes which form from the highest density peaks in the initial Gaussian random density field \citep[e.g.][and the references therein]{Mo2010}. Galaxies are therefore tracers of the underlying dark matter distribution, and with an accurate understanding of how biased these tracers are, the measurement of \textit{galaxy clustering} as a function of redshift and scale provides strong constraints on the properties of the $\Lambda$CDM model \citep[see for example][]{eboss2021}. It is becoming increasingly common to combine these two different `two-point' (2pt) statistics, along with a third measurement of the gravitational lensing of background galaxies by foreground galaxies, otherwise known as \textit{galaxy-galaxy lensing}.  These joint `3$\times$2pt' large-scale structure cosmological studies already have the precision to directly constrain some cosmological parameters independently of CMB measurements \citep{Heymans2020, DESY3}.

In this analysis we focus on exploiting the significant precision recovered with small-scale measurements of \textit{galaxy clustering} and \textit{galaxy-galaxy lensing}.  These non-linear scales are typically excluded from cosmological analyses owing to insufficient or uncertain modelling of the complex relationship between galaxies and the underlying matter distribution on these scales \citep{Davis1985, Dekel1987}. \textit{Galaxy bias} is scale-dependent, stochastic and changes as a function of galaxy luminosity, colour and morphological type \citep{Dekel1999, Zehavi2011, Cacciato2012a, Dvornik2018}. Based on these facts, it is not surprising that galaxy bias is generally considered a nuisance to be marginalised over in the recovery of cosmological constraints.  Many studies limit their analyses to scales where the galaxy bias is considered to be linear and scale independent \citep[see for example][]{VanUitert2018, Yoon2019, DESY3}. An alternative approach uses perturbation theory \citep{Desjacques2018}, to expand galaxy bias modelling into the mildly non-linear regime \citep[e.g.][]{Mandelbaum2013, Sanchez2017, Heymans2020, Pandey2021}.  However, as a measurement of small-scale galaxy bias also contains a wealth of information regarding galaxy formation, we argue that it is preferential to utilise all the data, along with an appropriate galaxy bias model, to facilitate joint constraints on both cosmology and galaxy bias.

Given our previous attempts to shine the light on the galaxy bias and its properties \citep{Dvornik2018}, in this analysis we adopt a realistic and physically motivated \textit{halo model} for galaxy bias. Under the assumption that all galaxies reside in dark matter haloes, we adopt a halo occupation distribution (HOD) model, a statistical description for how galaxies are distributed between and within the dark matter haloes \citep{Peacock2000, Scoccimarro2001, Mo2010, Yang2008, Cacciato2012, Bosch2012, Cacciato2013}. When combined with the halo model, which describes the non-linear matter distribution as a sum of spherical dark matter haloes \citep{Seljak2000, Cooray2002}, these models provide a fairly complete, broadly accurate\footnote{The standard halo model of dark matter haloes estimates the true power spectrum to within 20\% accuracy on non-linear scales \citep{Mead2015}.}, and easy to understand description of galaxy bias, halo masses, and galaxy clustering \citep{Cacciato2012}. 

Our approach builds on the cosmological analysis presented in \citet{Cacciato2012,More2014}, where the halo model is used to coherently analyse the clustering of galaxies, the galaxy-galaxy lensing signal \citep{Guzik2002, Yoo2006, Cacciato2009} and the galaxy abundances as a function of luminosity or stellar mass \citep{Bosch2012, Cacciato2012}. Furthermore, the same approach was used to study the galaxy-halo connection exclusively, with a fixed cosmology, by \citet{Leauthaud2011, Coupon2015}. This combination of probes is hereafter referred to as `2$\times$2pt+SMF', representing the combination of the two two-point statistics galaxy-galaxy lensing and galaxy clustering with the one-point stellar mass function. Analysis showing how much more information is in the 2$\times$2pt+SMF compared to the 2$\times$2pt is nicely presented in the paper from \citet{More2012}. The degeneracy breaking of model parameters arising from the inclusion of a stellar mass function is shown in  \citet{Mahony2022}.

Since early applications, there has been significant interest in using the halo model to interpret large-scale structure probes  \citet{Seljak2005,Cacciato2009,Li2009}.  The analysis of 2$\times$2pt statistics, down to non-linear scales, has been shown to lead to tight constraints for both $\Omega_{\mathrm{m}}$ and $\sigma_8$ \citep{Cacciato2012,Mandelbaum2013,More2014,Wibking2019}.  The halo model can also constrain extensions to the standard $\Lambda$CDM cosmologies, such as the equation of state of dark energy and neutrino mass \citep{More2012, Krause2016}. The choice of observables is motivated by the focus on a feasibility study on smaller scales which achieves similar precision, thus allowing for a direct and/or independent comparison to cosmic shear studies. In the era of high-precision cosmology, however, \citet{Miyatake2020} showed that the use of only a `broadly accurate' standard halo model leads to significant offsets in the recovered cosmological parameters from a 2$\times$2pt analysis of HOD-populated numerical simulations.  Consistency studies between the observed small-scale clustering and galaxy-galaxy lensing signals cast similar doubt on the accuracy of any standard halo model analysis \citep{Leauthaud2016,Lange2021,Amon2022}.  

Arguably the two most flawed approximations in the standard halo model formalism are that (i) haloes, and therefore galaxies, can overlap, and (ii) that haloes trace the matter distribution with a linear and scale independent halo bias. 
Previous attempts to improve these approximations have used halo exclusion formulations \citep{Giocoli2010} to solve the first problem, combined with radial bias functions which add scale dependence to the halo bias model \citep{Tinker2005, Bosch2012, Cacciato2012}.

In this analysis we instead follow the method proposed by \citet{Mead2021}, accounting for both scale-dependent non-linear halo bias and halo exclusion by incorporating the halo bias measured directly from the \textsc{DarkEmulator} suite of cosmological simulations \citep{Nishimichi2019}. As shown by \citet{Mahony2022}, this necessary upgrade to the standard halo model leads to sufficient accuracy in the recovered cosmological parameters from a 2$\times$2pt+SMF analysis, for the statistical power of current imaging surveys.

Other approximations in a halo model analysis include that the halo mass is the sole variable that determines the properties of the haloes and their occupying galaxies.  Galaxy properties and the clustering of haloes are, however, expected to have a secondary dependence, on their local environment and assembly history \citep[see][and references therein]{Wechsler2018}.  Furthermore the adopted halo density profile is modelled from dark matter-only numerical simulations, even though hydrodynamical simulations show that these profiles are modified by the presence of active galactic nuclei \citep{Schaller2015,Wang2020}. In \citet{Debackere2021}, it is shown that to account for baryon physics, it is sufficient to leave the concentration of dark matter haloes free. \citet{Amon2022} review the literature studies on the impact of these two approximations on 2$\times$2pt halo model studies with luminous red galaxies. Motivated by their conclusions, we choose to adopt nuisance parameters in our halo model to encapsulate the uncertainty of the impact of assembly bias and baryon feedback within our error budget.

In this paper, we analyse the most recent data release from the Kilo-Degree Survey \citep[KiDS-1000,][]{Kuijken2019,Giblin2020,Hildebrandt2020}, spanning over 1000 square degrees of imaging in nine bands from the optical through to the near-infrared.  Our main `KiDS-Bright' galaxy sample \citep{Bilicki2021} benefits from the 180 square degree overlap between KiDS and the spectroscopic Galaxy And Mass Assembly (GAMA) survey \citep{Driver2011}.  As an essentially complete spectroscopic survey to $r<19.8$, GAMA serves as an extensive training set for machine learning and the calibration of different sample selections. The resulting GAMA-trained photometric redshifts and stellar mass estimates for the KiDS-Bright sample have an enhanced accuracy and precision that benefits this galaxy-galaxy lensing and galaxy clustering study. In order to simultaneously constrain cosmology and galaxy bias, we use the 2$\times$2pt+SMF combination of galaxy clustering and galaxy-galaxy lensing, as well as constraints on galaxy abundances in the form of the stellar mass function. 

We improve previous related 2$\times$2pt+SMF studies by (i) using a more accurate analytical model with the addition of non-linear halo bias \citep{Mead2021}, taking into account the halo exclusion and scale dependence, (ii) taking the latest lensing and clustering data from a single survey, and (iii) using the full analytical covariance matrix including the cross-variance between all observables. Our analysis is highly complementary to the emulator based 2$\times$2pt halo model analysis of the Hyper Suprime Camera Survey \citep[HSC,][]{Miyatake2021}.

Throughout this paper, all radii and densities are given in comoving units, `$\log$' is used to refer to the 10-based logarithm, and `$\ln$' for natural logarithm. All the quantities that depend on the Hubble parameter adopt units of $h$, where $h = H_0 / 100$ km s$^{-1}$Mpc$^{-1}$. We also use $\overline{\rho}_{\text{m}}$ as the present day mean matter density of the Universe, \mbox{$\overline{\rho}_{\text{m}} = \Omega_{\text{m}, 0} \, \rho_{\text{crit}}$}, where \mbox{$\rho_{\text{crit}} = 3H^{2}_{0}/(8\pi G)$} and the halo masses are defined as $M = 4\pi r_{\Delta}^3 \Delta \; \overline{\rho}_{\text{m}} / 3 $ enclosed by the radius $r_\Delta$ within which the mean density of the halo is $\Delta$ times $\overline{\rho}_{\text{m}}$, with $\Delta = 200$.

This paper is organised as follows. In Sec. \ref{sec:halomodel_intro}, we review our analytical model to compute the galaxy stellar mass function, the galaxy-galaxy correlation function, and the galaxy-galaxy lensing signal using the halo model combined with a model that describes halo occupation statistics as a function of galaxy stellar mass. In Sec. \ref{sec:data}, we introduce the 2$\times$2pt+SMF KiDS measurements, specifics of the covariance calculation and our Bayesian analysis methodology. Our main results are presented in Sec. \ref{sec:results},  concluding our findings in Sec. \ref{sec:conclusions}.

\section{The Halo model}
\label{sec:halomodel_intro}

The halo model is an analytic framework that can be used to describe the clustering of matter and its evolution in the Universe \citep{Seljak2000, Peacock2000, Cooray2002, Bosch2012, Mead2015}. It is built upon the statistical description of the properties of dark matter haloes (namely the average density profile, large scale bias and abundance) as well as on the statistical description of the galaxies residing in them, using halo occupation distributions (HOD). The model is sufficiently flexible to consistently describe the statistical weak lensing signal around a selection of galaxies, their clustering, abundances and cosmic shear signal.

\subsection{Halo model ingredients}
\label{sec:halomodel_ingredients}

We assume that dark matter haloes are spherically symmetric on average, and have density profiles, $\rho(r \vert M) = M \, u_{\mathrm{h}}(r \vert M)$, that depend only on their mass $M$, and $u_{\mathrm{h}}(r \vert M)$ is the normalised density profile of a dark matter halo. Similarly, we assume that satellite galaxies in haloes of mass $M$ follow a  spherical number density distribution $n_{\mathrm{s}}(r \vert M) = N_{\mathrm{s}} \, u_{\mathrm{s}}(r \vert M)$, where $u_{\mathrm{s}}(r \vert M)$ is the normalised density profile of satellite galaxies. All central galaxies are positioned at the centre of their halo: $r=0$. We assume that the density profile of dark matter haloes follows a Navarro-Frenk-White (NFW) profile \citep{Navarro1997}. Since centrals and satellites are distributed differently, we write the galaxy-galaxy power spectrum, $P_{\mathrm{gg}}(k)$, as a combination of the central `c', satellite `s' and cross power spectrum, with
\begin{equation}\label{Pkgalsplit}
P_{\mathrm{gg}}(k) = A^{2}_{\mathrm{c}} P_{\mathrm{cc}}(k) + 
2 A_{\mathrm{c}} A_{\mathrm{s}} P_{\mathrm{cs}}(k) + A^{2}_{\mathrm{s}} P_{\mathrm{ss}}(k)\,,
\end{equation}
and the galaxy-matter power spectrum, $P_{\mathrm{gm}}(k)$,
\begin{equation}\label{Pkgaldmsplit}
P_{\mathrm{gm}}(k) = A_{\mathrm{c}} P_{\mathrm{cm}}(k) + A_{\mathrm{s}} P_{\mathrm{sm}}(k)\,.
\end{equation}
Here $A_{\mathrm{c}} = \overline{n}_{\mathrm{c}}/\overline{n}_{\mathrm{g}}$ and $A_{\mathrm{s}} =
\overline{n}_{\mathrm{s}}/\overline{n}_{\mathrm{g}} = 1 - A_{\mathrm{c}}$ are the central and satellite
fractions, respectively, and the average number densities $\overline{n}_{\mathrm{g}}$,
$\overline{n}_{\mathrm{c}}$ and $\overline{n}_{\mathrm{s}}$ follow from:
\begin{equation}\label{averng}
\overline{n}_{\mathrm{x}} = \int_{0}^{\infty} \langle N_{\mathrm{x}} \vert M \rangle \, n(M) \, \mathrm{d} M\,,
\end{equation}
where `x' stands for `g' (for galaxies), `c' (for centrals) or `s' (for satellites), $\langle N_{\mathrm{x}} \vert M \rangle$ is the average number of galaxies given halo mass $M$, and $n(M)$ is the halo mass function in the following form: 
\begin{equation}\label{eq:hmf}
n(M) = \frac{\overline{\rho}_{\mathrm{m}}}{M^{2}} \nu f(\nu) \frac{\mathrm{d} \ln \nu}{\mathrm{d} \ln M}\,,
\end{equation}
with $\nu = \delta_{\mathrm{c}} / \sigma(M)$ being the peak height. Here $\delta_{\mathrm{c}}$ is the critical overdensity required for spherical collapse at redshift $z$, and $\sigma(M)$ is the mass variance. For $f(\nu)$ we use the fitting function to the numerical simulations presented in \citet{Tinker2010}. In addition, it is common practice to split two-point statistics into a 1-halo term (both points are located in the same halo) and a 2-halo term (the two points are located in different haloes). The 1-halo terms are:
\begin{equation}
P^{\mathrm{1h}}_{\mathrm{cc}}(k) = {1 \over \overline{n}_{\mathrm{c}}}\,,
\end{equation}
\begin{equation}
P^{\mathrm{1h}}_{\mathrm{ss}}(k) = \mathcal{P} \int_{0}^{\infty} \mathcal{H}_{\mathrm{s}}^{2}(k, M) \, n(M) \, \mathrm{d} M\,,
\label{eqn:P1hss}
\end{equation}
and all other terms are given by:
\begin{equation}
P^{\mathrm{1h}}_{\mathrm{xy}}(k) = \int_{0}^{\infty} \mathcal{H}_{\mathrm{x}}(k, M) \, \mathcal{H}_{\mathrm{y}}(k, M) \, n(M) \, \mathrm{d} M\,.
\end{equation}
Here `x' and `y' are either `c' (central), `s' (satellite), or `m' (matter), $\mathcal{P}$ is a Poisson parameter that captures the scatter in the number of satellite galaxies at fixed halo mass (in this case a free parameter -- we define the $\mathcal{P}$ in detail using equations \ref{stoch_sat_poisson} and \ref{beta_def}) and we have defined the mass, central and satellite profiles as
\begin{equation}
\mathcal{H}_{\mathrm{m}}(k, M) = {M \over \overline{\rho}_{\mathrm{m}}} \,  \tilde{u}_{\mathrm{h}}(k \vert M)\,,
\end{equation}
\begin{equation}
\mathcal{H}_{\mathrm{c}}(k, M) = {\langle N_{\mathrm{c}} \vert M \rangle \over \overline{n}_{\mathrm{c}}} \,,
\end{equation}
and
\begin{equation}
\mathcal{H}_{\mathrm{s}}(k, M) = {\langle N_{\mathrm{s}} \vert M \rangle \over \overline{n}_{\mathrm{s}}} \,  \tilde{u}_{\mathrm{s}}(k \vert M)\,,
\end{equation}
with $\tilde{u}_{\mathrm{h}}(k \vert M)$ and $\tilde{u}_{\mathrm{s}}(k \vert M)$ the Fourier transforms of the  halo density profile and the satellite number density profile, respectively, both normalised to unity [$\tilde{u}(k \! \!= \!\! 0 \vert M) \!\! = \!\! 1 $]. The various 2-halo terms are given by:
\begin{align}\label{P2hcc}
P^{\mathrm{2h}}_{\mathrm{x}\mathrm{y}}(k) = P_{\mathrm{lin}}(k) \,  &\int_{0}^{\infty} \mathrm{d} M_1 \, \mathcal{H}_{\mathrm{x}}(k, M_1) \, b_{\mathrm{h}}(M_1)\, n(M_1) \nonumber \\\ 
&\times \int_{0}^{\infty} \mathrm{d} M_2 \, \mathcal{H}_{\mathrm{y}}(k, M_2) \, b_{\mathrm{h}}(M_2)\, n(M_2) \nonumber \\\
&+ P_{\mathrm{lin}}(k) \, I_{\mathrm{xy}}^{\mathrm{NL}}(k) \,
\end{align}
where $P_{\mathrm{lin}}(k)$ is the linear power spectrum, obtained using the \citet{Eisenstein1997} matter transfer function, and $b_\mathrm{h}(M,z)$ is the halo bias function. We adopt the \citet{Tinker2010} halo bias function which together with their halo mass function provides for consistent normalisation of the halo model integrals. The second term in equation \ref{P2hcc} encompasses the beyond-linear halo bias correction $\beta^{\mathrm{NL}}$ proposed by \citet{Mead2021} where, 
\begin{equation}
\begin{split}
I_{\mathrm{xy}}^{\mathrm{NL}}(k) & = \int_{0}^{\infty} \int_{0}^{\infty} \mathrm{d}M_1 \mathrm{d}M_2 \ \beta^{\mathrm{NL}}(k,M_1,M_2) \\
& \times \mathcal{H}_{\mathrm{x}}(k, M_1)\,\mathcal{H}_{\mathrm{y}}(k,M_2)\\
& \times n(M_1)\, n(M_2)\, b_{\mathrm{h}}(M_1)\, b_{\mathrm{h}}(M_2) \ .
\end{split}
\end{equation}
$\beta^{\mathrm{NL}}$ is measured using the \textsc{DarkQuest} emulator \citep{Nishimichi2019, Miyatake2020, Mahony2022}, by measuring the non-linear halo-halo power spectrum and then dividing it by the linear matter power spectrum multiplied with the product of linear bias factors \citep[][Eq. 23]{Mead2021}. Due to the definition of $\beta^{\mathrm{NL}}$, this measurement also holds true for galaxy-galaxy and galaxy-matter correlations. As shown in \citet{Mahony2022}, this function is cosmology dependent, but does not account for assembly bias effects. In this paper, owing to the volume-limited mix of all types of galaxies used in our analysis, we consider any assembly bias to be a subdominant effect as the secondary properties are unlikely to manifest for a non-specific galaxy type selection \citep{Wechsler2018}.
Numerically, the integrals in the halo model are not integrated from zero to infinity, but rather between a wide range of halo masses. Special care has to be taken to account for the masses outside of the integration limits, for which an appropriate correction is applied (as derived in \citet{Cacciato2009}, Eqs. 24 and 25, and in \citet{Mead2020, Mead2021}, Appendices A in both papers). The two-point correlation functions corresponding to these power-spectra are obtained by Fourier transformation:
\begin{equation}\label{xiFTfromPK}
\xi_{\mathrm{xy}}(r) = {1 \over 2 \pi^{2}} \int_{0}^{\infty} P_{\mathrm{xy}}(k) \, {\sin kr \over kr} \, k^{2} \, \mathrm{d} k\,. 
\end{equation}

For the halo bias function, $b_{\mathrm{h}}$, we use the fitting function from \citet{Tinker2010}, as it was obtained using the same numerical simulation from which the halo mass function was obtained. We have adopted the parametrisation of the concentration-mass relation, given by \citet{Duffy2011}:
\begin{equation}
\label{eq:con_duffy}
c(M, z) = 10.14\; \ \left[\frac{M}{(2\times 10^{12} M_{\odot}/h)}\right]^{- 0.081}\ (1+z)^{-1.01} \,.
\end{equation}
We allow for an additional normalisation $f_{\mathrm{h,s}}$, such that
\begin{equation}
c_{\mathrm{h,s}}(M, z) = f_{\mathrm{h,s}}\, c(M, z)\,,
\label{eqn:fsfh}
\end{equation} 
where $f_{\mathrm{h}}$ is the normalisation of the concentration-mass relation for dark matter haloes $\tilde{u}_{\mathrm{h}}(k \vert M)$, and $f_{\mathrm{s}}$ is the normalisation of the concentration-mass relation for the distribution of satellite galaxies $\tilde{u}_{\mathrm{s}}(k \vert M)$. The profiles $\tilde{u}_{\mathrm{h}}(k \vert M)$ and $\tilde{u}_{\mathrm{s}}(k \vert M)$ are both assumed to be non-truncated NFW profiles, with the same virial mass.
Our adoption here of separate concentration-mass relations for dark matter haloes and satellite galaxies provides enough flexibility in the model to capture the uncertain impact of baryon feedback \citep[for the scales adopted,][]{Debackere2020, Debackere2021, Amon2022}, and it has been used in the literature \citep{Cacciato2012, Viola2015, VanUitert2016, Dvornik2018} to account for such effects. This additional flexibility is motivated by the fact that in the simulations, the AGN feedback pushes the baryons and dark matter from halo centres towards outskirts, and by that effectively changing the concentration of the matter distribution \citep{Debackere2020, Mead2020}. 
This is also supported by observations \citep{Viola2015}, which showed that the preferred value for the concentration normalisation is lower than 1. Using the halo model with these extra parameters is a benefit over the emulators that are based on dark-matter only simulations \citep[as for instance the \textsc{DarkQuest} emulator][]{Nishimichi2019, Miyatake2020, Mahony2022}, since they do not offer a simple way to accommodate for such flexibility, nor require simulations \citep{Schneider2015}.

In the halo model we do not consider the mis-centred central term, as for a selection of galaxies the signature is accounted for through the terms for satellite galaxies, which do not reside in the centres of haloes by definition. What is more, the satellite galaxies populate haloes regardless of the existence of a central galaxy, which further removes the need for a mis-centred term (no central condition is enforced).

\subsection{Conditional stellar mass function}
\label{sec:halomodel_csmf}

We model the galaxy stellar mass function and halo occupation statistics using the Conditional Stellar Mass Function \citep[CSMF, motivated by][]{Yang2008a, Cacciato2009, Cacciato2012, Wang2013, VanUitert2016}. The CSMF, $\Phi(M_{\star} \vert M)$, specifies the average number of galaxies of stellar mass $M_{\star}$ that reside in a halo of mass $M$. In this formalism, the halo occupation statistics of central galaxies are defined via the function: 
\begin{equation}\label{CLFsplit}
\Phi(M_{\star} \vert M) = \Phi_{\mathrm{c}}(M_{\star}  \vert M) + \Phi_{\mathrm{s}}(M_{\star}  \vert M)\,.
\end{equation}
In particular, the CSMF of central galaxies is modelled as a log-normal,
\begin{equation}\label{phi_c}
\Phi_{\mathrm{c}}(M_{\star}  \vert M) = {1 \over {\sqrt{2\pi} \, {\ln}(10)\, \sigma_{\mathrm{c}} M_{\star} } 
}{\exp}\left[- { {\log(M_{\star} / M^{*}_{\mathrm{c}} )^2 } \over 2\,\sigma_{\mathrm{c}}^{2}} \right]\, \,,
\end{equation}
and the satellite term as a modified Schechter function,
\begin{equation}\label{phi_s}
\Phi_{\mathrm{s}}(M_{\star}  \vert M) = { \phi^{*}_{\mathrm{s}} \over M^{*}_{\mathrm{s}}}\,
\left({M_{\star} \over M^{*}_{\mathrm{s}}}\right)^{\alpha_{\mathrm{s}}} \,
{\exp} \left[- \left ({M_{\star} \over M^{*}_{\mathrm{s}}}\right )^2 \right] 
\,,
\end{equation}
where $\sigma_{\mathrm{c}}$ is the scatter between stellar mass and halo mass and $\alpha_{\mathrm{s}}$ governs the power law behaviour of satellite galaxies. Note that $M^{*}_{\mathrm{c}}$, $\sigma_{\mathrm{c}}$, $\phi^{*}_{\mathrm{s}}$, $\alpha_{\mathrm{s}}$ and
$M^{*}_{\mathrm{s}}$ are, in principle, all functions of halo mass $M$, but here we assume that $\sigma_{\mathrm{c}}$ and $\alpha_{\mathrm{s}}$ are independent of the halo mass $M$. Inspired by \citet{Yang2008a}, who studied the halo occupation properties of galaxies in the Sloan Digital Sky Survey, we parametrise $M^{*}_{\mathrm{c}}$, $M^{*}_{\mathrm{s}}$ and $\phi^{*}_{\mathrm{s}}$ as:
\begin{equation}\label{eq:CMF4}
M^{*}_{\mathrm{c}}(M) = M_{0} \frac{(M/M_{1})^{\gamma_{1}}}{[1 + (M/M_{1})]^{\gamma_{1} - \gamma_{2}}}\,,
\end{equation}
\begin{equation}\label{eq:CMF5}
M_{\mathrm{s}}^{*}(M) = 0.56\ M^{*}_{\mathrm{c}}(M)\,,
\end{equation}
and
\begin{equation}\label{eq:CMF7}
\log[\phi_{\mathrm{s}}^{*}(M)] = b_{0} + b_{1}(\log m_{13})\,,
\end{equation}
where $m_{13} = M/(10^{13}M_{\odot}h^{-1})$. In their analysis of the stellar-to-halo mass relation of GAMA galaxies \citet{VanUitert2016} find that varying the pre-factor of 0.56 in Equation \ref{eq:CMF5} does not significantly affect the results, therefore we retain this normalisation in our analysis. We can see that the stellar to halo mass relation for $M \ll M_{1}$ behaves as $M^{*}_{\mathrm{c}} \propto M^{\gamma_{1}}$ and for $M \gg M_{1}$, $M^{*}_{\mathrm{c}} \propto M^{\gamma_{2}}$, where $M_{1}$ is a characteristic mass scale and $M_{0}$ is a normalisation. Here $\gamma_{1}$, $\gamma_{2}$, $b_{0}$ and $b_{1}$ are all free parameters that govern the two slopes of the stellar-to-halo mass relation and the normalisation of the Schechter function. The choice of functional form of the CSMF is motivated by the good performance as seen in previous lensing and combined lensing and clustering studies. In Eq. \ref{eq:CMF4}, we adopt an effective stellar-to-halo mass relation for our mixed-population of red and blue galaxies. \citet{Bilicki2021} demonstrate a strong colour-dependence to this relationship, and future studies will investigate including a red/blue galaxy split in our analysis, which can also help to improve the modelling of intrinsic galaxy alignments \citep[e.g.][]{Li2021}.

From the CSMF it is straightforward to compute the galaxy stellar mass function (SMF) and the halo occupation numbers. The galaxy stellar mass function is in this case given by 
\begin{equation}\label{GSMF}
\Phi_{\mathrm{x}}(M_{\star}) = \int_{0}^{\infty} \Phi_{\mathrm{x}}(M_{\star} \vert M) \, n(M) \, \mathrm{d} M\,,
\end{equation}

and the average number of galaxies with stellar masses in the range $M_{\star,1} \leq M_{\star}  \leq M_{\star,2}$ is given by:
\begin{equation}\label{HODfromCLF}
\langle N_{\mathrm{x}} \vert M \rangle = \int_{M_{\star,1}}^{M_{\star,2}} \Phi_{\mathrm{x}}(M_{\star}  \vert M) \, \mathrm{d} M_{\star} \,,
\end{equation}

where `x' is either `c' (central), `s' (satellite), or the total contribution from all galaxies. 
In order to predict the satellite-satellite term for the galaxy clustering power spectra (Eq.~\ref{eqn:P1hss}), we use that 
\begin{equation}\label{stoch_sat_poisson}
\langle N_{\mathrm{s}}^{2} \vert M \rangle = \mathcal{P}(M) \langle N_{\mathrm{s}} \vert M \rangle^{2} + 
\langle N_{\mathrm{s}} \vert M \rangle \,,
\end{equation}
where $\mathcal{P}(M)$ is the mass-dependent Poisson parameter defined as: 
\begin{equation}\label{beta_def}
\mathcal{P}(M) \equiv  {\langle N_{\mathrm{s}}(N_{\mathrm{s}} - 1) \vert M \rangle \over  \langle N_{\mathrm{s}} \vert M \rangle^{2}} \,,
\end{equation}
which is unity if $\langle N_{\mathrm{s}} \vert M \rangle$ is given by a Poisson distribution, larger than unity if the distribution is wider than a Poisson distribution (also called super-Poissonian distribution) or smaller than unity if the distribution is narrower than a Poisson distribution (also called sub-Poissonian distribution). In our fiducial analysis we limit ourselves to cases in which $\mathcal{P}(M)$ is independent of halo mass, i.e., $\mathcal{P}(M) = \mathcal{P}$, and we treat $\mathcal{P}$ as a free parameter. In Appendix~\ref{sec:poisson_model} we present an extension to our fiducial analysis, allowing for mass-dependence in the Poisson parameter, based on the observed distribution of satellite galaxies in the GAMA group catalogue \citep{Robotham2011}. Our findings are sensitive to the selection criteria chosen for the GAMA group catalogue. We are nevertheless able to conclude that assuming the Poisson parameter is independent of halo mass impacts our primary cosmological parameter constraints (mostly $S_8$) at an acceptable $\sim 1\sigma$ level.

Overall, all the free parameters used to describe the halo occupation distributions and the connection with the dark matter are:
\begin{equation} 
\label{eq:hod_params}
\lambda^{\mathrm{HOD}} = [f_{\mathrm{h}}, M_0, M_1, \gamma_1, \gamma_2, \sigma_{\mathrm{c}}, f_{\mathrm{s}}, \alpha_{\mathrm{s}}, b_0, b_1, \mathcal{P}]\,.
\end{equation} 
Priors on these parameters are broad, assuming wide uniform distributions, similar to the priors used in two studies of the galaxy-halo connection that both used GAMA and KiDS data \citep{VanUitert2016, Dvornik2018}. The halo occupation distribution parameters could in principle also depend on redshift and halo mass, but furthering the complexity of the model, by increasing the number of parameters, would not be justified by the data. Our parameters describe an effective model over the redshift range in the analysis. Priors and their ranges can be found in Table \ref{tab:results}.

\subsection{Projected lensing and clustering functions}
\label{sec:halomodel_projected}

Once $P_{\mathrm{gg}}(k)$ and $P_{\mathrm{gm}}(k)$ have been determined, it is fairly straightforward to compute the projected galaxy-galaxy correlation function, $w_{\mathrm{p}}(r_{\mathrm{p}})$, and the excess surface density (ESD) profile, $\Delta\Sigma(r_{\mathrm{p}})$. The projected galaxy-galaxy correlation function, $w_{\mathrm{p}}(r_{\mathrm{p}})$, is related to the real-space galaxy-galaxy correlation function, $\xi_{\mathrm{gg}}(r)$, according to
\begin{align}\label{eq:wpRRSD}
w_{\mathrm{p}}(r_{\mathrm{p}})  &=  2 \int_{0}^{r_{\pi,\mathrm{max}}} \xi_{\mathrm{gg}}(r_{\mathrm{p}},r_{\pi},z)  \, \mathrm{d} r_{\pi} \nonumber \\
&=   2 \sum_{l=0}^2 \int_0^{r_{\pi,\mathrm{max}}} \xi_{2l}(s, z)\,\mathcal{L}_{2l}(r_{\pi}/s) \,\mathrm{d} r_{\pi} \,.
\end{align}
Here $\xi_{\mathrm{gg}}(r_{\mathrm{p}},r_{\pi},z)$ is the redshift-space galaxy-galaxy correlation function, $r_{\pi}$ is the redshift-space separation perpendicular to the line-of-sight and $r_{\pi,\mathrm{max}}$ is the maximum integration range used for the data (here we use $r_{\pi,\mathrm{max}} = 233 h^{-1} \mathrm{Mpc}$), $s = \sqrt{r^2_{\mathrm{p}} + r^2_{\pi}}$ is the separation between the galaxies, $\mathcal{L}_l(x)$ is the $l^{\mathrm{th}}$ Legendre polynomial, and $\xi_0$, $\xi_2$, and $\xi_4$ are given by
\begin{equation}\label{eq:monopole}
\xi_0(r,z) = \left( 1 + {2 \over 3}\beta_{\mathrm{k}} + {1 \over 5}\beta_{\mathrm{k}}^2\right) \,
\xi_{\mathrm{gg}}(r,z)\,,
\end{equation}
\begin{equation}\label{eq:quadrupole}
\xi_2(r,z) = \left( {4 \over 3}\beta_{\mathrm{k}} + {4 \over 7}\beta_{\mathrm{k}}^2\right) \,
\left[\xi_{\mathrm{gg}}(r,z) - 3 J_3(r,z)\right]\,,
\end{equation}
\begin{equation}\label{eq:octopole}
\xi_4(r,z) = {8 \over 35}\beta_{\mathrm{k}}^2 \, \left[\xi_{\mathrm{gg}}(r,z) + 
{15\over 2} J_3(r,z) - {35\over 2}J_5(r,z) \right]\,,
\end{equation}
where
\begin{equation}\label{eq:Jintegral}
J_n(r,z) = {1 \over r^n} \int_0^r \xi_{\mathrm{gg}}(y,z) \, y^{n-1} \, \mathrm{d} y\,. 
\end{equation}
and
\begin{equation}\label{eq:betapar}
\beta _{\mathrm{k}}= \beta_{\mathrm{k}}(z) = {1 \over \bar{b}(z)} 
\left({\mathrm{d} {\mathrm{ln}} D(z) \over \mathrm{d} {\mathrm{ln}} a}\right)_z
\end{equation}
with $a = 1/(1+z)$ the scale factor, $D(z)$ the linear growth factor, and 
\begin{equation}\label{avbias}
\bar{b}(z) = {1 \over \bar{n}_{\mathrm{g}}(z)} \int_{0}^{\infty} \langle N_{\mathrm{g}} \vert M \rangle \,
b_{\mathrm{h}}(M,z) \, n(M,z) \, \mathrm{d} M\,,
\end{equation}
the mean bias of the galaxies in consideration. Eq. \ref{eq:wpRRSD} accounts for the large-scale redshift-space distortions due to infall (the `Kaiser'-effect), which is necessary because the measurements for $w_{\mathrm{p}}(r_{\mathrm{p}})$ are obtained for a finite $r_{\mathrm{max}}$.  We note that whilst this \citet{Kaiser1987} formalism is only strictly valid in the linear regime, we adopt the non-linear galaxy-galaxy correlation function, $\xi_{\mathrm{gg}}(r)$, in Eqs. \ref{eq:monopole} - \ref{eq:octopole}, with the non-linearities captured through the halo model power spectra in Eq.~\ref{xiFTfromPK}.  \citet{Bosch2012} show that this modification provides a more accurate correction for the residual redshift space distortions, and that ignoring the presence of residual redshift space distortions leads to systematic errors that can easily exceed 20 percent on scales with $r_{\mathrm{p}} > 10 h^{-1}\, \mathrm{Mpc}$ \citep{Cacciato2012}.

The excess surface density profile (ESD), $\Delta \Sigma(r_{\mathrm{p}})$, is defined as
\begin{equation}\label{eq:shear} 
\Delta\Sigma(r_{\mathrm{p}}) = 
{2\over r_{\mathrm{p}}^2} \int_0^{r_{\mathrm{p}}} \Sigma(R') \, R' \, \mathrm{d} R'  - \Sigma(r_{\mathrm{p}}) \,.
\end{equation}
Here $\Sigma(r_{\mathrm{p}})$ is the projected surface mass density, which is
related to the galaxy-dark matter cross correlation, $\xi_{\mathrm{gm}}(r)$, according to
\begin{equation}\label{Sigma_approx}
\Sigma(r_{\mathrm{p}}) = 2\bar{\rho}_{\mathrm{m}} \int_{r_{\mathrm{p}}}^{\infty}  \xi_{\mathrm{gm}}(r) \, { r\, \mathrm{d} r \over \sqrt{r^2 - r_{\mathrm{p}}^2}}\,.
\end{equation}
The final model predictions and the covariance matrix are bin-averaged to the bin widths of the data vectors.

\subsection{Cosmological parameters}
\label{sec:cosmo_params}

The cosmological parameters in our model are described by the vector:
\begin{equation}
\label{eq:cosmo_model}
\lambda^{\mathrm{cosmo}} = [\Omega_{\mathrm{m}}, \sigma_8, h, n_s, \Omega_{\mathrm{b}}]\,.
\end{equation} 
As mentioned in Sect. \ref{sec:intro}, the goal of this paper is to use the ESD, $w_{\mathrm{p}}$ and SMF data to constrain $\sigma_8$ and $\Omega_{\mathrm{m}}$. Because of that, we set the priors for those two parameters to be uninformative and set their ranges following the latest KiDS cosmic shear analysis \citep{Asgari2020}. The last 3 cosmological parameters are shown to be poorly constrained using the ESD, $w_{\mathrm{p}}$, and SMF data \citep{Cacciato2012, Mandelbaum2013}, thus they form a set of secondary cosmological parameters with informative priors. Priors and their ranges can be found in Table \ref{tab:results}\footnote{Our prior range is larger than the range of available nodes in the \textsc{DarkQuest} emulator. Owing due to the iterative updates to the $\beta^{\mathrm{NL}}$ estimation and the quick convergence we find towards parameters within the emulator's range, this does not pose an issue.}. In Appendix \ref{sec:prior} we verify that our choice of priors do not inform the main cosmological parameters.

\section{Data and sample selection}
\label{sec:data}

In this analysis we combined three observables from the Kilo-Degree Survey (KiDS): galaxy abundances in the form of the galaxy stellar mass function, galaxy clustering in the form of the projected galaxy correlation function, and galaxy-galaxy lensing in the form of excess surface density profiles. Our KiDS observations were taken with OmegaCAM \citep{Kuijken2011}, a 268-million pixel CCD mosaic camera mounted on the VLT Survey Telescope. These instruments were designed to perform weak lensing measurements, with the camera and telescope combination providing a fairly uniform point spread function across the field-of-view \citep{deJong2013}. 

We analysed the latest data release of the KiDS survey \citep[KiDS-1000,][]{Kuijken2019}, containing observations from 1006 square-degree survey tiles. Specifics of the survey, the calibration of the source shapes and photometric redshifts are described in \citet{Kuijken2019, Giblin2020, Hildebrandt2020}, respectively. The companion VISTA-VIKING \citep{Edge2013} survey has provided complementary imaging in near-infrared bands (\textit{ZYJHK}$_{\mathrm{s}}$), resulting in a unique deep, wide, nine-band imaging dataset \citep{Wright2018}. The default photo-$z$ estimates provided as part of the KiDS survey were derived with the Bayesian Photometric Redshift approach \citep[\textsc{BPZ},][]{Benitez2000}.

We used shape measurements based on the \textit{r}-band images, which have an average seeing of $0.66$ arcsec. The galaxy shapes were measured using \emph{lens}fit \citep{Miller2013}, which has been calibrated using image simulations described in \citet{Kannawadi2018}. This provides galaxy ellipticities ($\epsilon_{1}$, $\epsilon_{2}$) with respect to an equatorial coordinate system, and an optimal weight.

The galaxies used for our lens and clustering sample were taken from the `KiDS-Bright' sample \citep{Bilicki2021}. This sample mimics the selection of GAMA galaxies \citep{Driver2011}, by applying the condition $m_{r} < 20.0$. For these galaxies a different method of determining the photometric redshifts was employed using the \textsc{ANNz2} (Artificial Neural Network) machine learning method \citep{Sadeh2016}, with the spectroscopic GAMA survey, which is 98.5\% complete to $r<19.8$, as a training set \citep{Bilicki2018, Bilicki2021}. Comparing the obtained redshifts with the spectroscopic redshifts from the matched galaxies between KiDS-Bright and GAMA, \citet{Bilicki2021} concluded that the \textsc{ANNz2} photo-z are highly accurate with a mean offset of $\delta_z =5 \times 10^{-4}$, and a scaled mean absolute deviation scatter of $\sigma_z =0.018 (1+z) $.

Stellar mass estimates for the KiDS-Bright sample are obtained using the \textsc{LePhare} template fitting code \citep{Arnouts1999, Ilbert2006}. In these fits, \textsc{ANNz2} photo-$z$ estimates are used as input redshifts for each source, treating them as if they were exact, neglecting the percent error associated with the \textsc{ANNz2} redshift. In practice, this error has little impact on the fidelity of the stellar mass estimates \citep{Taylor2011}. The estimates assume a \citet{Chabrier2003} initial mass function, the \citet{Calzetti1994} dust-extinction law, \citet{Bruzual2003} stellar population synthesis models, and exponentially declining star formation histories. The input photometry to \textsc{LePhare} is extinction corrected using the \citet{Schlegel1998} maps with the \citet{Schlafly2011} coefficients, as described in \citet{Kuijken2019}. 

\citet{Bilicki2021} found that the KiDS-Bright stellar mass estimates are in excellent agreement with independent stellar mass estimates from \citet{wright:2016} that combine GAMA spectroscopic redshifts with multi-wavelength imaging from 21 broadband filters from the far-UV to the far-IR.  The median offset is $M^\mathrm{KiDS}_{\star}/ M^\mathrm{GAMA}_{\star} = -0.09 \pm0.18$ dex.  \citet{Brouwer2021} estimated the overall systematic uncertainty on the stellar mass estimates of the KiDS-Bright sample, combining the uncertainty arising from the \textsc{LePhare} model fit, the photometric redshift scatter, and the difference found when exchanging elliptical aperture magnitudes for Sérsic model magnitudes.  They estimated an overall uncertainty of $\sigma_{M_*}=0.12$ dex for the KiDS-Bright sample.  This systematic uncertainty also includes the estimated Eddington systematic bias of $\sim 0.027$ dex \citep{Brouwer2021}, which is estimated from the population of red and blue galaxies and it is considered a worst-case scenario.  We choose to account for both statistical and systematic uncertainty in the stellar mass estimates through the nuisance parameter $\sigma_\mathrm{c}$, in equation~\ref{phi_c}, which provides the freedom to model both the intrinsic and measurement noise scatter in the stellar-to-halo mass relation \citep{Leauthaud2012, Bilicki2021}. Furthermore, as the systematic and statistical uncertainties are comparable in power, the entries in the SMF and cross-covariances are inflated by a factor of 2 to account for the uncertainty arising from Eddington bias and the systematic shift in stellar masses, and not only through the $\sigma_\mathrm{c}$ parameter. Due to the weak cosmology dependence of the SMF, this primarily increases only the uncertainty of our HOD parameters, as the SMF is in the first place used to break degeneracies in our HOD part of the halo model.

\subsection{Stellar mass function measurements and sample selection: SMF}
\label{sec:smf}

\begin{figure}
	\centering
	\includegraphics[width=\columnwidth]{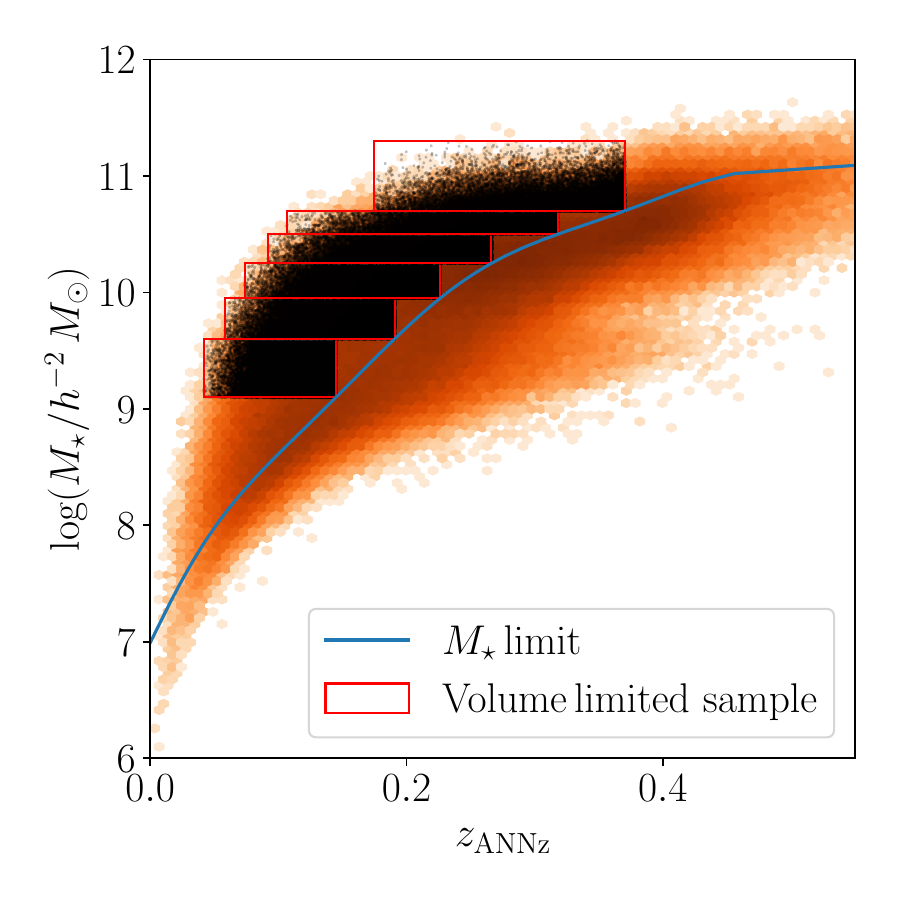}
 	\caption{Galaxy stellar mass as a function of \textsc{ANNz2} photometric redshift for the KiDS-Bright sample. The full sample is shown with a logarithmic hexagonal density plot. The blue line shows the stellar mass limit determined using the automated method presented by \citet{Wright2017}. Red boxes show the six stellar mass bins used in the analysis, with individual galaxies plotted as black dots. The bin ranges were chosen in such a way as to achieve a good signal-to-noise ratio in all bins for our galaxy-galaxy lensing and galaxy clustering measurements.}
	\label{fig:selection}
\end{figure}

Our SMF measurements are performed using the maximum-volume weighting method \citep{Schmidt1968, Saunders1990, Cole2011, Baldry2012, Wright2017}. We weight each galaxy $i$ by the inverse of the comoving volume over which the galaxy would be visible, given the magnitude limit of the whole sample, $1/V_{\mathrm{max},i}$. To estimate the number density $\Phi(M_{\star})$, we have to derive $M_{\star,\mathrm{lim}}(z)$, the completeness in stellar mass as a function of redshift for our flux-limited sample. For the $1/V_{\mathrm{max}}$ technique, we need to know $z_{\mathrm{max},i}$, the maximum redshift beyond which galaxy $i$ with stellar mass $M_{\star,i}$ would no longer be part of the subsample \citep{Weigel2016}. This is done by determining the point at which the sample begins to become incomplete. Usually this process contains a potentially biased visual inspection. To avoid any bias, we instead adopt the automated method presented by \citet{Wright2017}, using the \textsc{MassFuncFitR} package. The algorithm estimates the turn-over point of the number density distribution in bins of comoving distance and stellar mass independently. In each fine bin of comoving distance, we take the mass at the peak density as the mass turn-over point. In each fine bin of stellar mass, we take the largest comoving distance at median stellar mass density as the distance turn-over point. The obtained turn-over points are then fit with a high-degree polynomial resulting in a smooth form for the stellar mass limit as a function of redshift.  This limit can be compared to the $M_*-z_{\mathrm{ANNz2}}$ distribution of the full KiDS-Bright galaxies in Fig. \ref{fig:selection}.

\begin{figure}
	\centering
	\includegraphics[width=\columnwidth]{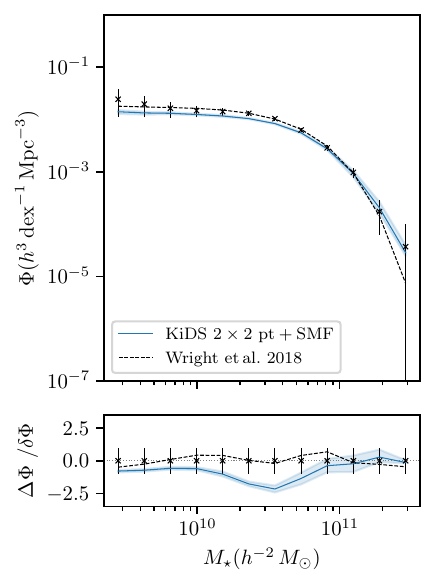}
 	\caption{The KiDS-Bright galaxy stellar mass function and the fractional errors. \textit{Upper panel:} The KiDS-Bright galaxy stellar mass function (crosses) compared to the model from \citet{Wright2018a}, evaluated at the median redshift of our sample (black dashed). The blue line and shaded region indicates the best fit model and 68\% confidence levels of our best fit halo model (Eq.~\ref{GSMF}). We caution that the quality of the fit cannot be judged by eye, because of the covariance in the data, between the data points and between the other observables. The reduced $\chi^2$ value for this observable is 1.05 ($\mathrm{DoF} = 14.58$, $p$-value = 0.39), estimated using the method presented in Appendix \ref{sec:fit}. \textit{Lower panel:} The fractional errors on the data and the model, $\Delta \Phi / \delta\Phi$.}
	\label{fig:smf}
\end{figure}

In Fig. \ref{fig:smf} we present the stellar mass function of the volume-limited KiDS-Bright sample from the galaxies in the 6 stellar mass bins, $\Phi(M_{\star})$, which has a median redshift of $z=0.25$.   This is determined from the galaxy counts within the stellar mass limit, with errors derived analytically in Appendix \ref{sec:smf_cov}. We find good agreement between the KiDS measurement and the stellar mass function from \citet{Wright2018a}, evaluated at the median redshift of our sample. \citet{Wright2018a} is based on from an analysis using spectroscopic data from GAMA, COSMOS and HST.  This comparison therefore demonstrates the agreement in the SMF between spectroscopic data and our photometric KiDS-Bright sample of galaxies, demonstrating that our stellar mass estimates are robust to the uncertainty in the photometric redshifts \citep{Taylor2011, Bilicki2021, Brouwer2021}.

As galaxy bias is inherently dependent on the stellar mass of the galaxy \citep{Dvornik2018}, we analyse the weak lensing and galaxy clustering of the KiDS-Bright galaxies grouped into 6 stellar mass bins.  We choose to limit our analysis to galaxies within the stellar mass range of $9.1 < \log(M_{\star} / h^{-2}\,M_{\odot}) \leq11.3$, with the number of bins, and bin limits chosen in such a way to achieve a similar and significant signal-to-noise ratio in all bins.  Using the redshift-dependent stellar mass limit, we define upper redshift bounds to ensure each stellar mass bin is volume-limited, as indicated with red boxes in Fig. \ref{fig:selection}.  The lower redshift bound is set to contain $95$ percent of the volume-limited sample.  The number of galaxies, median stellar mass and redshift of each bin is reported in Table~\ref{tab:sample}.

\begin{table}
	\caption{KiDS-Bright stellar mass samples: overview of the number of galaxies/lenses, median stellar masses $M_{\star, \mathrm{med}}$ and median redshifts $z_{\mathrm{med}}$. Stellar masses are given in units of $\log(M_{\star} / h^{-2}\,M_{\odot})$. The final sample of galaxies used is a small subsample of all KiDS-Bright galaxies ($\sim$ 1 million).}
	\centering
	\label{tab:sample}
	\begin{small}
	\begin{tabular}{lrrrrrrr} 
		\toprule
		Bin & \multicolumn{1}{c}{Range} & \multicolumn{1}{c}{$N_{\mathrm{gal}}$} & \multicolumn{1}{c}{$M_{\star, \mathrm{med}}$} & \multicolumn{1}{c}{$z_{\mathrm{med}}$} \\ 
		\midrule
		1 & (9.1, 9.6]    & 32\,846 & 9.33 & 0.12 \\
		2 & (9.6, 9.95]   & 35\,559 & 9.77 & 0.15 \\
		3 & (9.95, 10.25] & 39\,487 & 10.09 & 0.18 \\
		4 & (10.25, 10.5] & 38\,544 & 10.36 & 0.22 \\
		5 & (10.5, 10.7]  & 28\,814 & 10.58 & 0.27 \\
		6 & (10.7, 11.3]  & 22\,560 & 10.79 & 0.32 \\
		\bottomrule
	\end{tabular}
	\end{small} 
\end{table}

\subsection{Galaxy-galaxy lensing measurement: ESD}
\label{sec:gg_lensing}

As shown in \citet{Bilicki2021}, the excellent \textsc{ANNz2} photometric redshift estimates for the galaxies in the KiDS-Bright sample allow for robust estimates of their physical characteristics, in particular the stellar mass. In this section we combine this information with accurate shape measurements for more distant KiDS sources from \citet{Giblin2020} to measure the galaxy-galaxy lensing signal. To quantify the weak gravitational lensing signal we use source galaxies from KiDS DR4 with a \textsc{BPZ} photo-$z$ in the range $0.1<z_\mathrm{B}<1.2$. 

The lensing signal of an individual lens is too small to be detected, and hence we compute a weighted average of the tangential ellipticity $\epsilon_{\mathrm{t}}$ as a function of projected distance $r_{\mathrm{p}}$ using a large number of lens-source pairs. In the weak lensing regime this provides an unbiased estimate of the tangential shear, $\gamma_{\mathrm{t}}$, which in turn can be related to the excess surface density (ESD) $\Delta\Sigma(r_{\mathrm{p}})$, defined as the difference between the mean projected surface mass density inside a projected radius $r_{\mathrm{p}}$ and the mean surface density at $r_{\mathrm{p}}$ \citep[as in Eq. \ref{eq:shear}, for more details see appendix C from][]{Dvornik2018}.

We compute a weighted average to account for the variation in the precision of the shear estimate, captured by the \emph{lens}fit weight $w_{\text{s}}$ \citep[see][for details]{Conti2016,Kannawadi2018}, and the fact that the amplitude of the lensing signal depends on the source redshift. The weight assigned to each lens-source pair is
\begin{equation}
\label{eq:weights}
\widetilde{w}_{\mathrm{ls}}=w_{\mathrm{s}} \left(\widetilde \Sigma_{\mathrm{cr, ls}}^{-1}\right)^{2} \, ,
\end{equation}
the product of the \emph{lens}fit weight $w_{\text{s}}$ and the square of $\widetilde\Sigma_{\mathrm{cr, ls}}^{-1}$ -- the effective inverse critical surface mass density, which is a geometric term that downweights lens-source pairs that are close in redshift \mbox{\citep[e.g.][]{Bartelmann1999}}.

We compute the effective inverse critical surface mass density for each lens using the \mbox{photo-$z$} of the lens $z_{\mathrm{l}}$ and the full normalised redshift probability density of the sources, $n(z_{\mathrm{s}})$. The latter is calculated employing the self-organising map calibration method \citep{Wright2020} as applied to KiDS DR4 in \citet{Hildebrandt2020}. The resulting effective inverse critical surface density can be written as:
\begin{equation}
\label{eq:crit_effective}
\widetilde\Sigma_{\mathrm{cr, ls}}^{-1}=\frac{4\pi G}{c^2} \int_{0}^{\infty} (1+ z_{\mathrm{l}})^{2}  D(z_{\mathrm{l}}) \left(\int_{z_{\mathrm{l}}}^{\infty} \frac{D(z_{\mathrm{l}},z_{\mathrm{s}})}{D(z_{\mathrm{s}})}n(z_{\mathrm{s}}) \, \mathrm{d}z_{\mathrm{s}} \right) \, p(z_{\mathrm{l}}) \, \mathrm{d}z_{\mathrm{l}} \,,
\end{equation}
where $D(z_{\mathrm{l}})$, $D(z_{\mathrm{s}})$, $D(z_{\mathrm{l}},z_{\mathrm{s}})$ are the angular diameter distances to the lens, source, and between the lens and the source, respectively. For the lens redshifts \mbox{$z_{\mathrm{l}}$} we use the \textsc{ANNz2} \mbox{photo-$z$} of the KiDS-Bright foreground galaxy sample. We implement the contribution of $z_{\mathrm{l}}$ by integrating over the redshift probability distributions $p(z_{\mathrm{l}})$ of each lens. The lensing kernel is wide and therefore the resulting ESD signals are not sensitive to the small wings of the lens redshift probability distributions. We can thus safely approximate $p(z_{\mathrm{l}})$ as a normal distribution centred at the lenses photo-$z$, with a standard deviation $\sigma_{\mathrm{z}} / (1+z_{\mathrm{l}}) = 0.018$ \citep{Bilicki2021}. From previous KiDS GGL studies we know that the error on the mean and width of source $n(z_{\mathrm{s}})$ are not biasing the galaxy-galaxy lensing signal \citep[as shown in][]{Dvornik2017}.

For the source redshifts $z_{\mathrm{s}}$ we follow the method used in  \citet{Dvornik2018}, by integrating over the part of the redshift probability distribution $n(z_{\mathrm{s}})$ where $z_{\mathrm{s}} > z_{\mathrm{l}}$. The galaxy source sample is specific to each lens redshift $z_{\mathrm{l}}$, with a minimum photometric redshift $z_{\mathrm{s}} = z_{\mathrm{l}} + \delta_{z}$, with $\delta_{z} = 0.2$ that is used to remove sources that are physically associated with the lenses. Thus, the ESD can be directly computed in bins of projected distance $r_{\text{p}}$ to the lenses as:
\begin{equation}
\label{eq:ESDmeasured}
\Delta \Sigma_{\text{gm}} (r_{\mathrm{p}}) = \left[ \frac{\sum_{\mathrm{ls}}\widetilde{w}_{\mathrm{ls}}\epsilon_{\mathrm{t, s}}\Sigma_{\mathrm{cr, ls}}^{\prime}}{\sum_{\mathrm{ls}}\widetilde{w}_{\mathrm{ls}}} \right] \frac{1}{1+\overline{m}} \, .
\end{equation}
where $\Sigma_{\mathrm{cr, ls}}^{\prime} \equiv 1/ \widetilde\Sigma_{\mathrm{cr, ls}}^{-1}$, the sum is over all source-lens pairs in the distance bin, and
\begin{equation}
\overline{m} = \frac{\sum_{i}w_{i}^{\prime}m_{i}}{\sum_{i}w_{i}^{\prime}} \, ,
\end{equation}
is an average correction to the ESD profile that has to be applied to account for the multiplicative bias $m$ in the \emph{lens}fit shear estimates. 
The sum goes over thin redshift slices for which $m_{i}$ is obtained using image simulations \citep{Kannawadi2018}, weighted 
by $w^{\prime} = w_{\mathrm{s}}\,D(z_{\mathrm{l}},z_{\mathrm{s}}) / D(z_{\mathrm{s}})$ for a given lens-source sample. The value of $\overline{m}$ is $-0.003$ for the 6 stellar mass bins, independent of the scale at which it is computed. The uncertainty in $m$ is not marginalised over, as the contribution of the central $m$ value is at most a percent of the total error budget of the galaxy-galaxy lensing signal.

\begin{figure}
	\centering
	\includegraphics[width=\columnwidth]{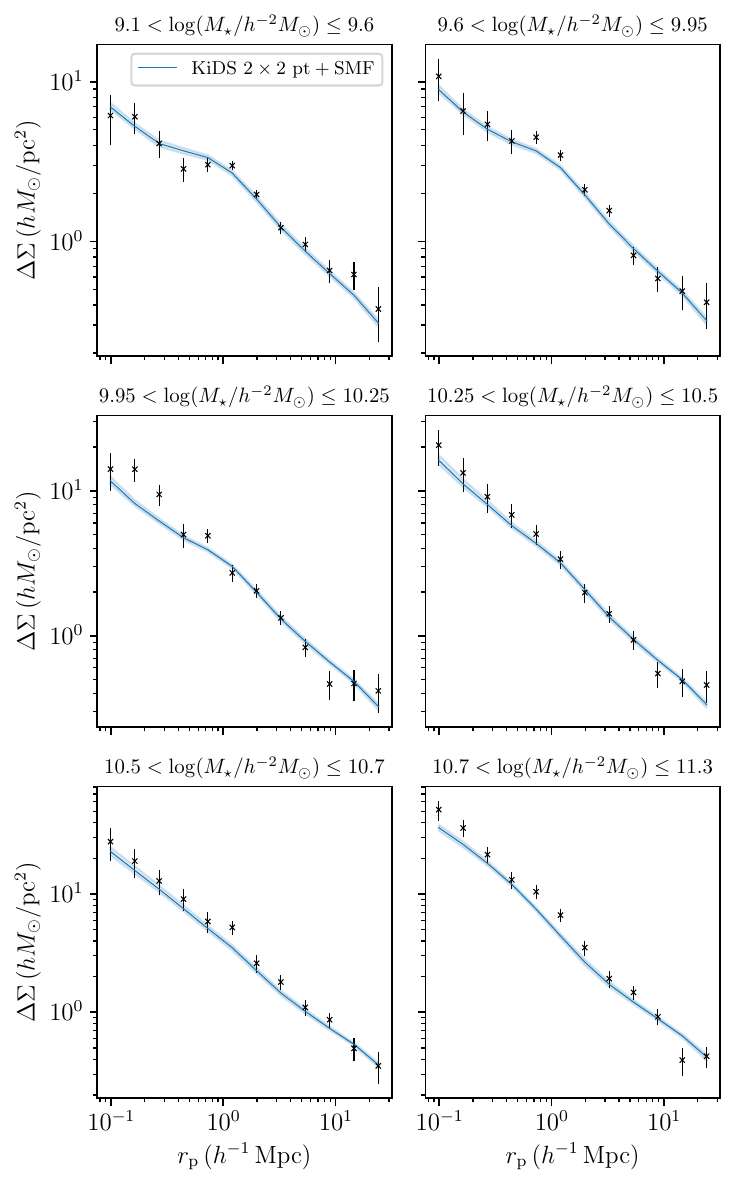}
 	\caption{Galaxy-galaxy lensing: The stacked ESD profiles of the six stellar mass bins in the KiDS-Bright galaxy sample defined in Table \ref{tab:sample}. The solid lines represent the best-fitting fiducial ESD halo model (Sec. \ref{sec:halomodel_projected}, Eq.~\ref{eq:shear}) as obtained using an MCMC fit, with the 68 percent confidence interval indicated with a shaded region. We caution that the quality of the fit cannot be judged by eye, because of the covariance in the data between the observed bins and also between the observables. The reduced $\chi^2$ value for this observable is 1.28 ($\mathrm{DoF} = 73.18$, $p$-value = 0.05), estimated using the method presented in Appendix \ref{sec:fit}.}
	\label{fig:esd}
\end{figure}

We note that  the measurements presented here are not corrected for the contamination of the source sample by galaxies that are physically associated with the lenses (the so-called `boost correction'). The impact on $\Delta\Sigma$ is minimal, because of the weighting with the inverse square of the critical surface density in Eq.~\eqref{eq:crit_effective}, \citep[see for instance the bottom panel of fig.~A4 in][]{Dvornik2017} and the removal of the sources physically associated with the lens from our signal measurements. The effect of using photometric lenses in the ESD measurements is directly accounted for in our estimator and the covariance matrix. We subtract the signal around random points, which suppresses any large-scale systematics and sample variance \citep{Singh2016}.  This empirical `random' correction for large-scale sample variance has been shown to improve robustness on the measurement scales which are particularly relevant to constrain linear bias \citep{Dvornik2018}. We find the random correction for the KiDS-Bright sample becomes significant at scales $R \gtrsim 3 h^{-1}\, \mathrm{Mpc}$, rising to more than 100\% of the ESD signal in the three lowest stellar mass bins, and it thus dictates the range of measurement scales we use in the analysis. On these large scales the random correction is more than four times larger than the statistical uncertainty (see Appendix~\ref{sec:randoms} for details). The resulting random-corrected galaxy-galaxy lensing ESD measurements for the six stellar mass bins are shown in Fig. \ref{fig:esd}

\subsection{Projected galaxy clustering measurements: $w_{\mathrm{p}}$}
\label{sec:proj_clustering}

We measure the clustering of the KiDS-Bright galaxy sample using the Landy-Szalay \citep{Landy1993} estimator for the galaxy correlation function
\begin{equation}
    \hat{\xi}_{\mathrm{gg}}(r_{\mathrm{p}},r_{\pi}) = \frac{DD - 2DR + RR}{RR} \Bigg{|}_{r_{\mathrm{p}},r_{\pi}} \, .
\end{equation}
Here we count the number of galaxy-galaxy ($DD$), random-random ($RR$) and galaxy-random ($DR$) pairs, as a function of the pair's transverse $r_{\mathrm{p}}$ and radial $r_{\pi}$ comoving separation. The accuracy of galaxy clustering measurements with this estimator depends critically on the quality of the random, $R$, catalogues. We use the \citet{Johnston2020} organised random methodology that has been shown to recover unbiased clustering measurements in a series of mock galaxy catalogue analyses for the KiDS-Bright sample.  Using machine learning, we infer the high-dimensional mapping between the observed on-sky galaxy number density and three systematic-tracer variables; atmospheric seeing, point spread function ellipticity and limiting magnitude.  Systematically-induced density variations across the survey footprint can then be defined.  We randomly distribute clones of the real galaxies across the survey footprint, preserving the on-sky systematic density patterns, and matching the on-sky systematic-tracer properties to that of the clone's parent galaxy.  By retaining the photometric properties of the parent for each clone, selection effects are accurately mirrored in the organised randoms for any galaxy sub-sample, for example the 6 different stellar mass bins in our analysis. We use 20 times more randoms than data points as presented by \citet{Johnston2020}.

The projected clustering correlation function is estimated through an integral over the line-of-sight separation, limited by a maximum defined distance $r_{\pi, \mathrm{max}}$,
\begin{equation}
    \hat{w}_{\mathrm{p}}(r_{\mathrm{p}}) = \int^{r_{\pi, \mathrm{max}}}_{-r_{\pi, \mathrm{max}}} \hat{\xi}_{\mathrm{gg}}(r_{\mathrm{p}},r_{\pi}) \, \mathrm{d}r_{\pi} \, .
    \label{pau:eq:wgg_xiggintegrated}
\end{equation}
When analysing spectroscopic data, this continuous integral is estimated using a discrete sum, typically adopting uniform bins in $r_{\pi}$, with $r_{\pi, \mathrm{max}}$ ranging from $40 h^{-1}\mathrm{Mpc}$ to $100 h^{-1}\mathrm{Mpc}$ \citep[as in for instance][]{Mandelbaum2010, Farrow2015}. Here the $r_{\pi, \mathrm{max}}$ limits are chosen to maximise the number of correlated galaxy pairs along the line-of-sight in the presence of redshift space distortions, whilst minimising the noise arising from the inclusion of uncorrelated objects.  With our KiDS-Bright photometric sample we have an additional uncertainty in the true redshift, $\sigma_z = 0.018(1+z)$, which translates into an uncertainty on the radial distance of the order $\sim 100 h^{-1}\mathrm{Mpc}$,  This renders the approach taken for spectroscopic samples sub-optimal in terms of signal-to-noise.  We therefore choose to follow the approach of \citet{Johnston2021} who optimised the projected galaxy clustering analysis of the photometric Physics of the Accelerating Universe Survey (PAUS), using dynamic binning in $r_{\pi}$ out to a maximum $r_{\pi} = 233 h^{-1}\mathrm{Mpc}$. This is motivated by the fact that PAUS photometric redshifts show a similar uncertainty as the KiDS-Bright sample. Using a mock galaxy catalogue, \citet{Johnston2021} demonstrated that by allowing for an increase in the bin size from small to large values of $r_{\pi}$, their approach maximises the count of physically associated objects, whilst minimising noise at large-$r_{\pi}$ with the broader bin size.  Given the similar photometric redshift properties of KiDS-Bright and PAUS, we adopt their 12-$r_{\pi}$-bin adapted Fibonacci sequence in our estimator.

\begin{figure}
	\centering
	\includegraphics[width=\columnwidth]{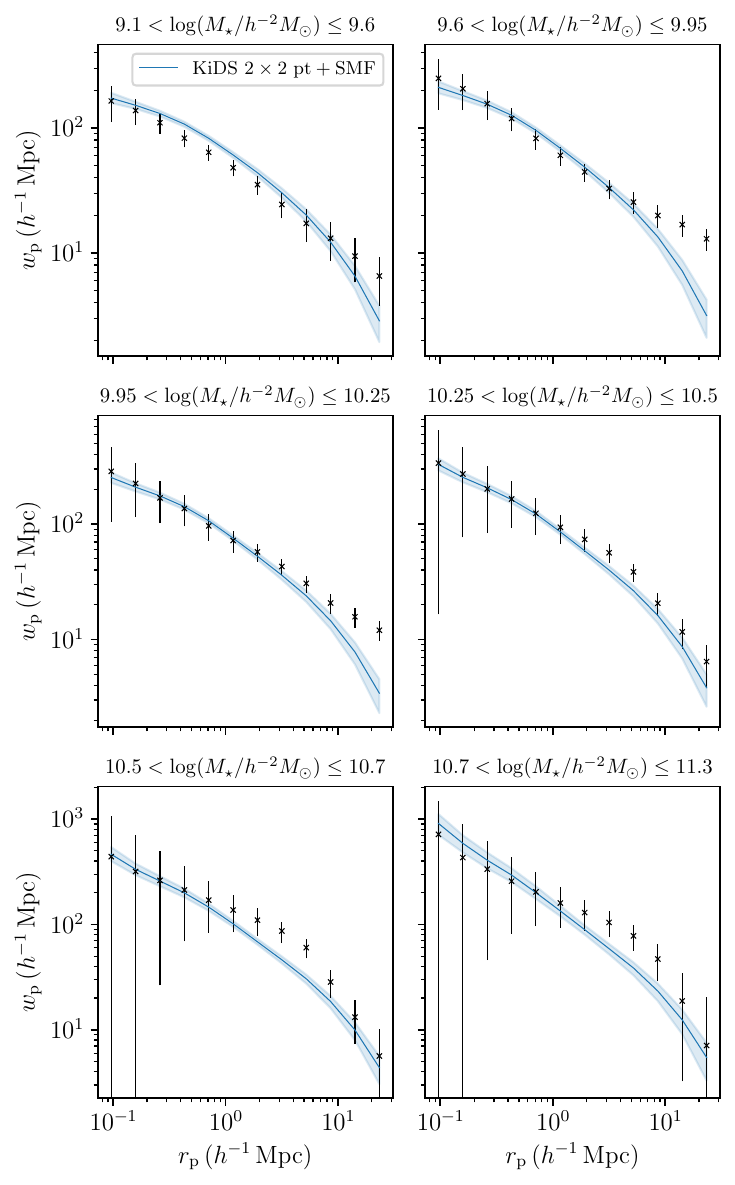}
 	\caption{Galaxy clustering: The projected galaxy clustering signal of the six stellar mass bins in the KiDS-Bright galaxy sample defined in Table \ref{tab:sample}. The solid lines represent the best-fitting fiducial halo model (Sec. \ref{sec:halomodel_projected}, Eq.~\ref{eq:wpRRSD}) as obtained using an MCMC fit, with the 68 percent confidence interval indicated with a shaded region. We caution that the quality of the fit cannot be judged by eye, because of the covariance in the data, between the observed bins and between the observables. The reduced $\chi^2$ value for this observable is 1.42 ($\mathrm{DoF} = 71.62$, $p$-value = 0.01), estimated using the method presented in Appendix \ref{sec:fit}.}
	\label{fig:wp}
\end{figure}

\citet{Johnston2021} analysed mock GAMA galaxy catalogues with PAU-like photometric redshifts to compare the projected clustering correlation function estimator $\hat{w}_{\mathrm{p}}(r_{\mathrm{p}})$ with the measurements using spectroscopic redshifts.  Adopting dynamic binning and random galaxy catalogues that mimic both the position and photometric redshift uncertainty of the real galaxy sample, they found a roughly scale-independent bias with $\hat{w}_{\mathrm{p}}/{w}_{\mathrm{p}^{\mathrm{spec}}} \simeq 0.8$.   As such the dynamic binning and organised randoms only partially correct the correlation functions for the dilution introduced by photometric redshift uncertainty.  Future work will focus on accounting for this dilution effect accurately in the theoretical prediction.  For the purposes of this analysis, however, we choose to include a free dilution parameter $\mathcal{D}$, which is used to correct the galaxy clustering measurements in the following way:
\begin{equation}
     \hat{w}_{\mathrm{p, corr}}(r_{\mathrm{p}}) = [1+\mathcal{D}]\,  \hat{w}_{\mathrm{p}}(r_{\mathrm{p}})\,.
     \label{eqn:Dfact}
\end{equation}
We adopt a uniform prior for $\mathcal{D}$ with the range between 0 and 0.3 and use a single parameter to scale all six stellar mass bins.  This prior was motivated by a series of mock KiDS-Bright galaxy clustering analysis using MICE2 \citep{Fosalba2015a, Fosalba2015, Crocce2015, Carretero2015, Hoffmann2015}, where we confirmed the findings of \citet{Johnston2021} and found no strong dependence of the dilution effect on stellar mass.  We note that a similar correction was applied to the Dark Energy Survey (DES) photometric clustering measurements \citep[][referred therein as $X_{\mathrm{lens}}$]{Pandey2021,DESY3}.  The prior and motivation behind the introduction of their systematic nuisance parameter differs, however. The resulting projected clustering measurements for the six stellar mass bins are shown in Fig. \ref{fig:wp}.

\subsection{Accounting for the cosmology dependence of distance measures}

To obtain estimates of the stellar mass function (SMF, Sect. \ref{sec:smf}, Fig. \ref{fig:smf}), the galaxy-galaxy lensing (ESD , Sect. \ref{sec:gg_lensing}, Fig. \ref{fig:esd}), and the projected galaxy clustering ($w_{\mathrm{p}}$, Sect. \ref{sec:proj_clustering}, Fig. \ref{fig:wp}), we have adopted a fiducial flat $\Lambda$CDM cosmology with $\Omega_{\mathrm{m}} = 0.3$ to compute distances.   As such our 2$\times$2pt+SMF data vector is cosmology-dependent, with changes in the fiducial cosmology changing the distance-redshift relation, which in turn shifts galaxies between the stellar mass bins and lens-source pairs between the radial separation bins.

At the mean redshift of the KiDS-Bright sample, the effect of changing $\Omega_{\mathrm{m}}$ within our prior limits introduces changes in distance estimates at the level of few percent. The approximation that the measurements are effectively independent of cosmological parameters within their observational uncertainties \citep{Mandelbaum2013, Cacciato2012} no longer holds for surveys with a statistical power that is similar or better than KiDS. 

In this analysis we account for the cosmology dependence of our data vector following the correction procedure presented in \citet{More2013} and \citet{More2014}, which modifies the model prediction for each cosmology targeted by the likelihood sampler.
 
First we define a cosmology-dependent comoving separation $r_{\mathrm{p}}^{\mathrm{model}}$ for our target model, relative to the comoving separation $r_{\mathrm{p}}$ that has been used to calculate our data vectors at a fixed fiducial cosmological model,
\begin{equation}
    r_{\mathrm{p}}^{\mathrm{model}} = r_{\mathrm{p}}^{\mathrm{fid}} \left[ \frac{\chi(z_{\mathrm{med}}, \mathcal{C}^{\mathrm{model}})}{\chi(z_{\mathrm{med}}, \mathcal{C}^{\mathrm{fid}})} \right] \,.
\end{equation}
Here $\chi$ is the comoving distance to the median lens redshift $z_{\mathrm{med}}$ in our target cosmological model $\mathcal{C}^{\mathrm{model}}$, or in our fiducial cosmological model $\mathcal{C}^{\mathrm{fid}}$. The galaxy clustering prediction for our target model is then given by
\begin{equation}
    \widetilde{w}_{\mathrm{p}}(r_{\mathrm{p}}) = w_{\mathrm{p}}(r_{\mathrm{p}}^{\mathrm{model}}) \left[ \frac{E^{\mathrm{model}}(z_{\mathrm{med}})}{E^{\mathrm{fid}}(z_{\mathrm{med}})} \right] \,,
\end{equation}
where $E(z)$ is the Hubble parameter. The galaxy-galaxy lensing prediction for our target model is given by
\begin{equation}
    \widetilde{\Delta\Sigma}(r_{\mathrm{p}}) = \Delta\Sigma(r_{\mathrm{p}}^{\mathrm{model}}) \left[ \frac{\Sigma_{\mathrm{cr}}^{\mathrm{model}}(z_{\mathrm{med}}, z_{\mathrm{s}})}{\Sigma_{\mathrm{cr}}^{\mathrm{fid}}(z_{\mathrm{med}}, z_{\mathrm{s}})} \right] \,,
\end{equation}
where $\Sigma_{\mathrm{cr}}$ is the critical surface density calculated for the median redshift of the lenses $z_{\mathrm{med}}$ and a fixed source redshift $z_{\mathrm{s}}= 0.6$.  Note that calculating the more precise estimate for $\Sigma_{\mathrm{cr}}$ using Eq.~\ref{eq:crit_effective} is not necessary in this instance, as $\Sigma_{\mathrm{cr}}$ only has a weak cosmology dependence. Finally the predictions of abundances of galaxies in the target cosmology is given by
\begin{equation}
    \widetilde{n}_{\mathrm{g}} = \overline{n}_{\mathrm{g}}^{\mathrm{model}} \left[ \frac{\chi^{3}(z_{\mathrm{u}}, \mathcal{C}^{\mathrm{model}}) - \chi^{3}(z_{\mathrm{l}}, \mathcal{C}^{\mathrm{model}})}{\chi^{3}(z_{\mathrm{u}}, \mathcal{C}^{\mathrm{fid}}) - \chi^{3}(z_{\mathrm{l}}, \mathcal{C}^{\mathrm{fid}})} \right] \,,
\end{equation}
which is implicitly correcting the surveyed volume in the stellar mass function calculation. Here the $z_{\mathrm{l}}$ and $z_{\mathrm{u}}$ are the lower and upper redshift limits in our samples.

\subsection{Covariance matrix}
\label{sec:covariance}

The covariance matrix used in this analysis is based on the analytical approach detailed in \citet{Dvornik2018} and \citet{Joachimi2020}, with the addition of the analytical covariance matrix for the SMF and the cross terms between the SMF and 2-point correlation functions. The new terms for the SMF covariance and the cross covariance between the SMF and 2-point functions are presented in Appendix \ref{sec:smf_cov}. Our implementation of the analytical covariance derivation was validated against theory \citep{Pielorz2010, Takada2013, Li2014, Marian2015, Krause2016}, independent software by \citet{Joachimi2020} and simulations \citep[MICE2][]{Fosalba2015a, Fosalba2015, Carretero2015, Crocce2015, Hoffmann2015}, following the validation approach of \citet{Blake2020} and \citet{Joachimi2020}. Survey area effects on the variance were calculated using the accurate, survey dependent and data based \textsc{Healpix} method presented in \citet{Joachimi2020}, equation E.10.

\subsection{Likelihood and iterative updates}
\label{sec:likely}

We use Bayesian inference to determine the posterior probability distribution $P(\boldsymbol{\theta}\, \vert\, \mathbf{d})$ of the model parameters $\boldsymbol{\theta}$, given the data $\mathbf{d}$. According to Bayes' theorem, $P(\boldsymbol{\theta}\, \vert\, \mathbf{d})$ is:
\begin{equation}
P(\boldsymbol{\theta}\, \vert\, \mathbf{d})= {P(\mathbf{d}\, \vert\, \boldsymbol{\theta}) P(\boldsymbol{\theta}) 
\over P(\mathbf{d})} \,,
\end{equation}
where $P(\mathbf{d}\, \vert\, \boldsymbol{\theta})$ is the likelihood of the data given the model parameters, $P(\boldsymbol{\theta})$ is the prior
probability of these parameters, and
\begin{equation}
P(\mathbf{d})= \int P(\mathbf{d} \vert\, \boldsymbol{\theta}) \, P(\boldsymbol{\theta}) \,  \mathrm{d} \boldsymbol{\theta}
\end{equation}
is the evidence for the model. Since, we do not perform model selection in this analysis, the
evidence just acts as a normalisation constant which we do not need to calculate. Given this, the likelihood distribution $P(\mathbf{d}\, \vert\, \boldsymbol{\theta})$ is assumed to be Gaussian:
\begin{equation}
P(\mathbf{d}\, \vert\, \boldsymbol{\theta}) = \frac{1}{\sqrt{(2\pi)^{n}\vert \mathbf{C} \vert}} \exp \left[- \frac{1}{2}\left[\left(\mathbf{m} (\boldsymbol{\theta}) - \mathbf{d} \right)^{T}{\mathbf{C}}^{-1} \left(\mathbf{m} (\boldsymbol{\theta}) - \mathbf{d}\right) \right] \right], 
\end{equation}
where $\mathbf{C}$ is the full covariance matrix for all the observables, containing their auto- and cross-correlations, $\vert \mathbf{C} \vert$ its determinant, $\mathbf{m} (\boldsymbol{\theta})$ the model given the parameters $\boldsymbol{\theta}$, and $n$ the number of observable bins. 
Priors can be found in Table \ref{tab:results}. For the Bayesian inference we use the MCMC sampler \textsc{ emcee} \citep{Foreman-Mackey2012}.

The posterior distribution in such highly multi-dimensional parameter spaces has numerous degeneracies and can be very difficult to sample from. Thus the choice of proposal distributions is very important in order to achieve fast convergence and reasonable acceptance fractions for the proposed walker positions. To do so, we combine the default stretch move in the \texttt{emcee} with the proposal function based on the kernel density estimator of the complementary ensemble of walkers \citep{Foreman-Mackey2012}\footnote{The moves are further defined in the documentation of the \texttt{emcee} package at \url{https://emcee.readthedocs.io}.} in such way that at every step of the sampler run, there is a 50\% chance to use one of the proposal methods. This setup has one downside, and that is that it uses many walkers, and thus computing power. On the other hand, the convergence is faster and the resulting auto-correlation times are shorter, giving us shorter MCMC chains overall.

During the MCMC runs we iteratively update the $\beta^{\mathrm{NL}}$ measurement (as it is cosmology dependent), as running the emulator at each step of the chain is computationally not feasible. Thus the $\beta^{\mathrm{NL}}$ measurement is evaluated using the median of the current position of the walkers in the parameter space. This returns an effective value for the non-linear halo bias correction that is, over the run of the MCMC, representative of the median of corrections that would be applied to every single model iteration in the chain. In our pipeline, the number of steps between iterations can be set by the user and we find that updating the $\beta^{\mathrm{NL}}$ values every 20 steps allows for a reasonable run time while providing enough updates to the $\beta^{\mathrm{NL}}$ correction. On the other hand, the covariance matrix is only re-evaluated with the new parameters at the end of the MCMC run and checked. We find that the updated covariance matrix and halo model parameters do not affect the results of our fit as our initial cosmological and HOD parameters were set to the one of \citet{Heymans2020} and \citet{VanUitert2016} which are close to our final results. The covariance matrix is dominated by the shot and/or shape noise on the majority of scales.

To report our result we use two methods to estimate our constraints and parameter values. One method uses the maximum statistics of the marginal posterior distributions for each parameter (MMAX). Here the asymmetric errors are estimated around the maximum point in iso-distribution levels to cover $68\%$ of the marginal distribution. For the second method we use the full posterior distribution to find the best-fitting parameters (maximum a posteriori point, MAP) and use the methodology presented in \cite{Joachimi2020} to associate an error to this measurement with the projected highest posterior density (PJ-HPD) approach. While the former method produces more stable parameter errors, especially when the likelihood surface is sparsely sampled, its point estimates, in general, do not correspond to the best fitting parameter values. In contrast the latter method will in general produce noisier error estimates with unbiased parameter values.

\section{Results}
\label{sec:results} 

\begin{table}
    \fontsize{8.5pt}{10.25pt}\selectfont
    \centering 
    \caption{Marginal constraints on all model parameters listed together with their priors.}
    \label{tab:results}
    \setlength\tabcolsep{4pt}
    \begin{tabular}{cccccc}
        \toprule
        Parameter & Prior & \multicolumn{2}{c}{Fiducial} \\ 
        & & MMAX & MAP+PJ-HPD \\
        \midrule
        $\Omega_{\mathrm{m}}$ & $[0.1, 0.6]$ & $0.290^{+0.021}_{-0.017}$ & $0.307^{+0.002}_{-0.031}$ \\  \addlinespace
        $\sigma_{8}$ & $[0.4, 1.2]$ & $0.781^{+0.033}_{-0.029}$ & $0.801^{+0.013}_{-0.041}$ \\ \addlinespace
        $h$ & $[0.64, 0.82]$ & $<0.726$ & $<0.711$ \\ \addlinespace
        $\Omega_{\mathrm{b}}$ & $[0.01, 0.06]$ & $>0.01$ & $>0.01$ \\ \addlinespace
        $n_{s}$ & $[0.92, 1.1]$ & $<1.004$ & $<0.978$ \\
        \midrule
        $S_8$ & -- & $0.773^{+0.028}_{-0.030}$ & $0.809^{+0.001}_{-0.055}$ \\
        \midrule
        $f_{\mathrm{h}}$ & $[0.0, 1.0]$ & $>0.645$ & $>0.939$ \\ \addlinespace
        $M_0$ & $[7.0, 13.0]$ & $10.519^{+0.039}_{-0.062}$ & $10.521^{+0.062}_{-0.044}$ \\ \addlinespace
        $M_1$ & $[9.0, 14.0]$ & $11.138^{+0.099}_{-0.132}$ & $11.145^{+0.136}_{-0.098}$ \\ \addlinespace
        $\gamma_1$ & $[2.5, 15.0]$ & $7.096^{+2.144}_{-1.406}$ & $7.385^{+1.441}_{-1.824}$ \\ \addlinespace
        $\gamma_2$ & $[0.0, 10.0]$ & $0.201 {\pm 0.010}$ & $0.201 {\pm 0.009}$ \\ \addlinespace
        $\sigma_{\mathrm{c}}$ & $[0.0, 2.0]$ & $0.108^{+0.067}_{-0.011}$ & $0.159^{+0.007}_{-0.070}$ \\ \addlinespace
        $f_{\mathrm{s}}$ & $[0.0, 1.0]$ & $>0.377$ & $>0.84$ \\ \addlinespace
        $\alpha_{\mathrm{s}}$ & $[-5.0, 5.0]$ & $-0.858^{+0.048}_{-0.052}$ & $-0.847^{+0.013}_{-0.097}$ \\ \addlinespace
        $b_0$ & $[-5.0, 5.0]$ & $-0.024^{+0.108}_{-0.117}$ & $-0.120^{+0.199}_{-0.001}$ \\ \addlinespace
        $b_1$ & $[-5.0, 5.0]$ & $1.149^{+0.091}_{-0.081}$ & $1.177^{+0.058}_{-0.096}$ \\ \addlinespace
        $\mathcal{P}$ & $[0.0, 2.0]$ & $0.403 {\pm 0.029}$ & $0.417^{+0.024}_{-0.013}$ \\ \addlinespace
        $\mathcal{D}$ & $[0.0, 0.3]$ & $0.144^{+0.091}_{-0.085}$ & $0.051^{+0.172}_{-0.006}$ \\
        \midrule
        $\chi^2_{\mathrm{red}}$ & -- & \multicolumn{2}{c}{1.07} \\ \addlinespace
        $p$-value & -- & \multicolumn{2}{c}{0.27} \\ \addlinespace

        \bottomrule
        \addlinespace
    \end{tabular}
    
    \caption*{Notes: This table lists all the free parameters in our model:  the energy density of cold matter $\Omega_{\mathrm{m}}$, the normalisation of power spectrum  $\sigma_8$, the dimensionless Hubble parameter $h$, the spectral index $n_s$, the energy density of baryonic matter $\Omega_{\mathrm{b}}$, the derived parameter $S_8$, the normalisation of the concentration-mass relation for dark matter haloes $f_{\mathrm{h}}$, the normalisation of stellar-to-halo mass relation $M_0$, the characteristic scale of the stellar-to-halo mass relation $M_1$, the slope parameters of the stellar-to-halo mass relation $\gamma_1$ and $\gamma_2$, the scatter between stellar mass and halo mass $\sigma_{\mathrm{c}}$, the normalisation of the concentration-mass relation for distribution of satellite galaxies $f_{\mathrm{s}}$, the power law behaviour of satellites $\alpha_{\mathrm{s}}$, the normalisation constants of the Schechter function $b_0$ and $b_1$, and the Poisson parameter $\mathcal{P}$. Parameters are deemed unconstrained when the marginal probability at $2\sigma$ level exceeds 13\% of the peak probability \citep[see appendix A of][]{Asgari2020}. In cases where one side is constrained we report the $1\sigma$ lower/upper limit. The MMAX estimate is the marginal maximum statistic, reporting the point of maximum marginal posterior distribution to the iso-posterior levels above and below the maximal point. The MAP+PJ-HPD (maximum posterior with projected joint highest posterior density) estimates are calculated following \citet{Joachimi2020}.
    }
\end{table}

Now we turn the focus to our results, presenting cosmological parameter constraints in Sect.~\ref{sec:fid_results}, large-scale analysis in Sect.~\ref{sec:large}, constraints on the galaxy-halo connection in Sect.~\ref{sec:fid_hod}, and effect of modelling of satellite galaxies in Sect~\ref{sec:poisson_model}.  Further details are presented in Appendix \ref{sec:tests}. To recap, our theoretical 2$\times$2pt+SMF model consists of 17 free parameters, two of which are our main cosmology parameters, with 3 more secondary cosmology parameters that are harder to constrain given the combination of observables, and 11 parameters describing the galaxy-halo connection in the form of the CSMF. With six stellar mass bins, and three observables, our combined data vector consists of 156 data points.  In Appendix \ref{sec:fit} we use mock data realisations to estimate the effective number of degrees of freedom for our analysis finding $\nu_{\mathrm{eff}}=147.55$.  Following the likelihood analysis described in Sect.~\ref{sec:likely} we are able to constrain 12 parameters, listed in Table~\ref{tab:results} along with their prior ranges.  We find the MAP to provide a good fit\footnote{We define an acceptable fit when $p(\chi^2 | \nu_{\mathrm{eff}}) \geq 0.003$, corresponding to less than a $3\sigma$ event, \citep[see the discussion in][]{Heymans2020}.  We note that \citet{DESY3} define a more stringent requirement where $p(\chi^2 | \nu_{\mathrm{eff}}) \geq 0.01$.  We find the goodness of fit for our 2$\times$2pt+SMF analysis, and each individual component of the data vector, to be acceptable given both these definitions.} to the data, with a reduced $\chi^2$ value of 1.07 and $p(\chi^2 | \nu_{\mathrm{eff}})  = 0.27$.

We compare the prediction from our fiducial model, and its 68\% confidence regions, to the measured galaxy abundance SMF in Fig.~\ref{fig:smf}, the galaxy-galaxy lensing ESD in Fig.~\ref{fig:esd}, and the galaxy clustering $w_{\mathrm{p}}$ in Fig.~\ref{fig:wp}, for all of the six stellar mass bins.  We find that the model reproduces the overall trends in the data, such as the presence of the bump at $\sim 1 h^{-1}\, \mathrm{Mpc}$ in ESD due to satellite galaxies, and the fact that the stronger signal is present where galaxies have higher stellar mass, showing that massive galaxies reside in more massive haloes.  We note that some caution is needed when interpreting the results, as the quality of the fit cannot be judged by eye due to highly correlated data points. We find acceptable fits to each component of our 2$\times$2pt+SMF data vector (for details see Appendix~\ref{sec:fit}).   We note that the poorest fit is found for the $w_{\mathrm{p}}$ section of our joint data vector with $p[\chi^2(w_{\mathrm{p}}) | \nu_{\mathrm{eff}}^{w_{\mathrm{p}}}]  = 0.01$.  Whilst a formally acceptable fit, this may indicate that our model is lacking the ability to correctly describe the photometric redshift dilution effect discussed in Sect.~\ref{sec:proj_clustering}.

\subsection{Cosmology constraints}
\label{sec:fid_results}

\begin{figure*}
	\centering
	\includegraphics[width=0.75\textwidth]{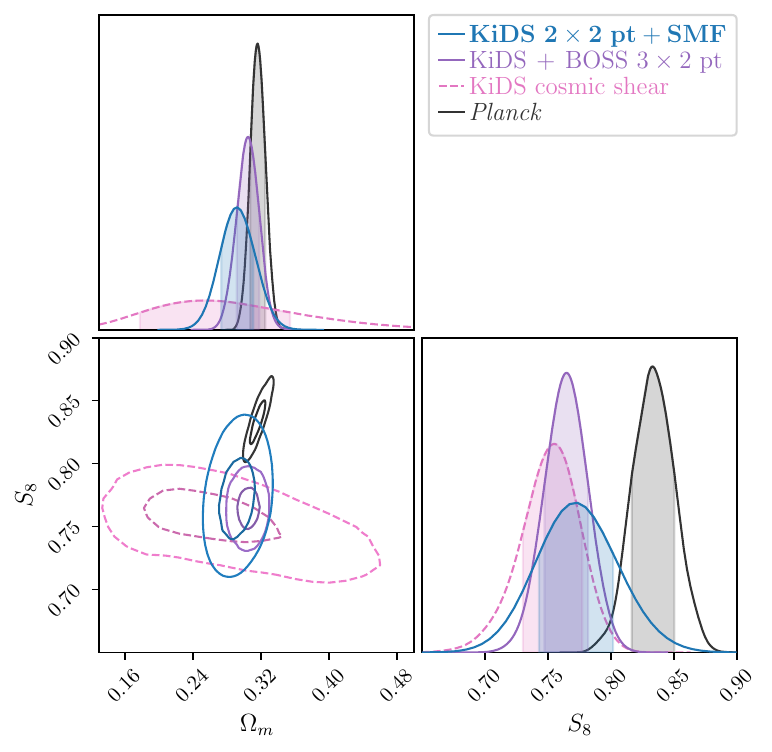}
 	\caption{Marginalised constraints for the joint distributions of $S_8$ and $\Omega_{\mathrm{m}}$. The 68\% and 95\% credible regions for the 2$\times$2pt+SMF fiducial analysis (blue) can be compared with constraints from KiDS cosmic shear \citep[][pink]{Asgari2020}, KiDS with BOSS 3$\times$2pt \citep[][purple]{Heymans2020}, and the CMB \citet[][black]{PlanckCollaboration2018}.}
	\label{fig:s8_corner}
\end{figure*}

\begin{figure*}
	\centering
	\includegraphics[width=0.9\textwidth]{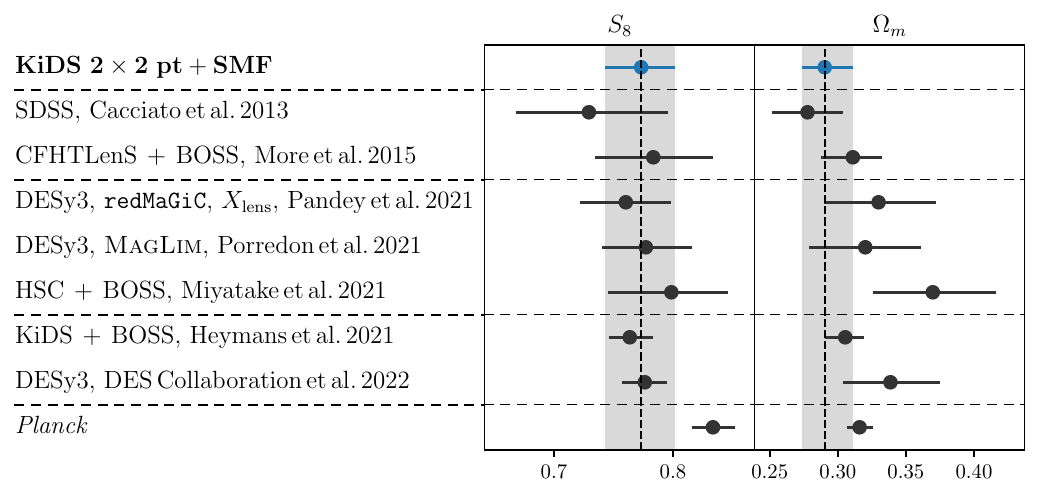}
 	\caption{A joint-probe comparison of $S_8$ and $\Omega_{\mathrm{m}}$ constraints. Our fiducial results (KiDS 2$\times$2pt +SMF) can be compared to a similar 2$\times$2pt +SMF analysis from SDSS \citep{Cacciato2012}; a series of 2$\times$2pt studies including the latest results from DES \citep{Porredon2021, Pandey2021} and HSC \citep{Miyatake2021}; 3$\times$2pt analyses from KiDS with BOSS \citep{Heymans2020} and DES \citep{DESY3}. The last entry shows the \citet[][TT,TE,EE+lowE]{PlanckCollaboration2018} constraints.  Our results are consistent with all studies, including \citet{PlanckCollaboration2018}, although we find a mild tension between our $S_8$ constraints and those from \textit{Planck}, at the level of $1.9\sigma$.}
	\label{fig:s8_summary}
\end{figure*}

We find the following cosmological parameter constraints from our simultaneous 2$\times$2pt+SMF analysis of the ESD, $w_{\mathrm{p}}$ and SMF signals of galaxies in the KiDS-Bright sample, 
\begin{align}
    \Omega_{\mathrm{m}} &= 0.290^{+0.021}_{-0.017} \nonumber \\
    \sigma_8 &= 0.781^{+0.033}_{-0.029} \nonumber \\ 
    S_8 &= 0.773^{+0.028}_{-0.030}\,, \nonumber
\end{align}
where $S_8 = \sigma_{8} \sqrt{\Omega_{\mathrm{m}}/0.3}$, and we quote the maximum statistics of the marginal posterior distributions (MMAX).  The remaining cosmological parameters are unconstrained by our analysis, and informed by our choice of prior (see Fig.~\ref{fig:full_corner}).  In Fig.~\ref{fig:s8_corner} we present the 68\% and 95\% confidence levels of the joint two-dimensional, marginalised posterior distribution in the $S_8 - \Omega_{\mathrm{m}}$ plane.  The 2$\times$2pt+SMF constraints are shown to be in good agreement with constraints from KiDS cosmic shear and KiDS with BOSS 3$\times$2pt constraints \citep{Asgari2020, Heymans2020}. They are formally consistent, but in some mild tension with the \citet{PlanckCollaboration2018} TT,TE,EE+lowE CMB results. Using the Hellinger distance as a tension measure \citep[see][]{Heymans2020}, the mild tension between our fiducial results and \textit{Planck} is $1.9\sigma$ in $S_8$.

In Fig. \ref{fig:s8_summary} we compare our constraints to a broader range of joint-probe large-scale structure analyses, finding consistency with all. \citet{Cacciato2012} and \citet{More2014} adopt a similar methodology to this study, using a halo model formalism to jointly analyse galaxy-galaxy lensing, galaxy clustering and galaxy abundance observables. \citet{Cacciato2012} analysed the Sloan Digital Sky Survey (SDSS-DR7). \citet{More2014} combined data from the Baryon Oscillation Spectroscopic Survey (BOSS) and the Canada France Hawaii Telescope Lensing Survey (CFHTLenS).  The improved constraining power in this KiDS analysis reflects the significant increase in depth for the lensing sample relative to SDSS-DR7, and the ten-fold increase in area relative to CFHTLenS.   We can also compare to 2$\times$2pt analyses that combine galaxy-galaxy lensing and galaxy clustering that introduce conservative scale cuts to reflect known limitations their adopted galaxy bias models. These include the fiducial analyses from DES, which adopt a linear galaxy bias model\footnote{\citet{Porredon2021, Pandey2021} also explore small-scale 2$\times$2pt analyses using non-linear galaxy bias models, finding a 20-30\% gain in cosmological constraining power.} \citep{Porredon2021, Pandey2021}, and from the Hyper Suprime Camera survey \citep[HSC,][]{Miyatake2021}.   The HSC analysis is highly complementary to our study as both analyses used a form of halo model with a halo occupation distribution. \citet{Miyatake2021} use the \citet{Zheng2005a} HOD built into the dark-matter N-body \textsc{DarkQuest} emulator to predict the 2$\times$2pt observables  \citep{Nishimichi2019, Miyatake2020}.   We use the same emulator to calibrate the non-linear halo bias in our model (see Sect.~\ref{sec:halomodel_ingredients}).  Compared to this analysis, HSC adopts more conservative scale cuts arising from concern over unmodeled baryon feedback on small-scales, which our methodology can account for through the free normalisation of the mass-concentration relation(see Sect.~\ref{sec:fid_hod}).  \citet{Miyatake2021} also choose scale cuts to mitigate small-scale assembly bias for the relatively rare luminous red galaxies in their sample, which cannot be modelled using an HOD. Owing to the volume-limited mix of all galaxies used in our KiDS-Bright analysis, we consider any assembly bias to be a subdominant effect in our theoretical model.

The constraining power of the KiDS 2$\times$2pt+SMF analysis in the $S_8-\Omega_{\mathrm{m}}$ plane is the same as that of the 3$\times$2pt studies from \citet{DESY3}, with $\sigma^{S_8}\sigma^{\Omega_{\mathrm{m}}}_{[\mathrm{DES\, 3 \times 2pt}]} / \sigma^{S_8}\sigma^{\Omega_{\mathrm{m}}}_{[\mathrm{KiDS\, 2\times 2pt+SMF}]} = 0.97$. This may be surprising given the five-fold increase in area for DES relative to KiDS, and the addition of the cosmic shear probe in the 3$\times$2pt analysis.  This comparison therefore highlights the significant constraining power from the inclusion of non-linear scales in the 2$\times$2pt+SMF analysis that are excluded from the DES 3$\times$2pt analysis.  
Comparing KiDS 2$\times$2pt+SMF constraints to the KiDS with BOSS 3$\times$2pt analysis \citep{Heymans2020}, we first review the BOSS spectroscopic clustering constraints where $\sigma^{S_8}\sigma^{\Omega_{\mathrm{m}}}_{[\mathrm{BOSS\, 1\times 2pt}]} / \sigma^{S_8}\sigma^{\Omega_{\mathrm{m}}}_{[\mathrm{KiDS\, 2\times 2pt+SMF}]} = 1.03$.  Finding the same constraining power between these analyses may again be surprising, given the nine-fold increase in area for BOSS relative to KiDS.  As such it demonstrates the significant constraining power from non-linear scales when the galaxy bias can be constrained using galaxy-galaxy lensing and galaxy abundance.  Comparing to the full 3$\times$2pt analysis we find $\sigma^{S_8}\sigma^{\Omega_{\mathrm{m}}}_{[\mathrm{KiDS+BOSS\, 3\times 2pt}]} / \sigma^{S_8}\sigma^{\Omega_{\mathrm{m}}}_{[\mathrm{KiDS\, 2\times 2pt+SMF}]} = 0.39$, where the extra constraining power in the 3$\times$2pt analysis is driven by the cosmic shear.  Future studies with KiDS will combine 2$\times$2pt+SMF with cosmic shear data, including further development and validation of our adopted halo model methodology.

In Appendix~\ref{sec:tests} we explore a number of extensions to our fiducial analysis.  In Appendix~\ref{sec:ia_mag_test} we quantify the expected contamination to our observables from intrinsic galaxy alignments and magnification.  The contamination levels are found to be negligible relative to our statistical errors, justifying our choice to not account for these astrophysical effects in our model.    In Appendix~\ref{sec:ns_model} we demonstrate that our $S_8$ and $\Omega_{\mathrm{m}}$ constraints are insensitive to our choice of prior on $n_s$.  In Appendix~\ref{sec:dilution} we quantify the bias in $\Omega_{\mathrm{m}}$ without the inclusion of our nuisance parameter $\mathcal{D}$ to model photometric redshift dilution in our galaxy clustering measurement. In Appendix~\ref{sec:poisson_model} we quantify the impact of assumptions governing the behaviour of satellite galaxies.

\subsection{Large-Scale Analysis}
\label{sec:large}

\citet{Amon2022} present a detailed analysis of the uncertainty on the amplitude of the small-scale galaxy-galaxy lensing and galaxy clustering signal for BOSS galaxies that arises from our imperfect knowledge of baryon feedback and assembly bias.  They conclude that the introduction of scale cuts with $r_{\mathrm{p}} > 5 h^{-1}\, \mathrm{Mpc}$ fully isolates these effects. Following the methodology in Appendix~\ref{sec:fit} we determine $\nu_{\mathrm{eff}}^{r_{\mathrm{p}}>5} = 60.31$ for a large-scale only 2$\times$2pt+SMF data vector analysis, calculating the large-scale goodness of fit of our fiducial best-fit model (see Table~\ref{tab:results}) to be $p[\chi^2 (r_{\mathrm{p}} > 5) | \nu_{\mathrm{eff}}^{r_{\mathrm{p}}>5}] = 0.19$.  As such we find no sign of tension between our fiducial all-scale analysis and a restricted large-scale analysis.  
 
The majority of the information in our $(r_{\mathrm{p}}>5)$ data vector comes from the SMF as there are only 4 highly correlated data points remaining in the ESD and $w_{\mathrm{p}}$ measurements, per stellar mass bin.  We found that a full likelihood analysis of the $(r_{\mathrm{p}}>5)$ data vector was unable to converge in our highly-flexible 17-parameter model space. Where larger scales are used exclusively, either more precise data or the use of a less flexible model is necessary \citep[such as in][amongst others]{More2013, Miyatake2021, Amon2022}.  The extra flexibility afforded in our model is, however, essential when analysing small scales in order to capture baryonic effects \citep{Debackere2020, Debackere2021}

\subsection{Galaxy-halo connection}
\label{sec:fid_hod}

\begin{figure}
	\centering
	\includegraphics[width=\columnwidth]{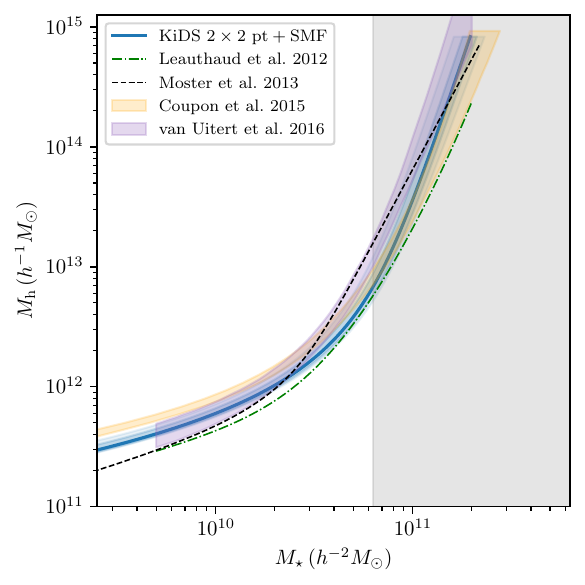}
 	\caption{The stellar-to-halo mass relation, as defined by Eq. \ref{eq:CMF4} using the best-fit HOD parameters from our 2$\times$2pt+SMF analysis (blue).  The result can be compared to results from \citet[][green]{Leauthaud2012}, \citet[][black]{Moster2013}, \citet[][orange]{Coupon2015}, and \citet[][purple]{VanUitert2016}. The grey area shows the range in stellar masses where the obtained stellar-to-halo mass relation is extrapolated beyond the range of median stellar masses used in this analysis.}
	\label{fig:smhm}
\end{figure}

The powerful aspect of the method used in this work is that it is able to simultaneously constrain both cosmological parameters as well as the halo occupation statistics. The full results are listed in Table \ref{tab:results} and the marginalised posterior distributions of all the parameters are shown in Fig. \ref{fig:full_corner}. The HOD parameters are tightly constrained, with some strong degeneracies between the parameters $M_{0}$, $M_{1}$ and $\gamma_{1}$ that govern the characteristic mass of the stellar-to-halo mass relation knee and the high mass slope of centrals, and between the $b_{0}$ and $b_{1}$ parameters, which govern the normalisation of the satellite CSMF. The first degeneracy is somewhat expected given the data, as the stellar mass function at the high mass end is highly uncertain and dominated by the Eddington bias. The second degeneracy also arises from the fact that both parameters compete for the overall normalisation of the satellite CSMF. 

We find the parameters of our HOD model to be in good agreement with the previous studies using GAMA-like galaxies \citep{VanUitert2016, Dvornik2018, Bilicki2021}. In order to show the agreement in a more intuitive way, we take the parameters of the stellar-to-halo mass relation and combine them using the same functional form (Eq. \ref{eq:CMF4}), which results in the relation shown in Fig. \ref{fig:smhm}. In the same figure we show good agreement with the results from \citet{VanUitert2016}, where the HOD parameters were constrained using galaxy-galaxy lensing combined with a SMF for KiDS and GAMA data, adopting a fixed {\it Planck} cosmology. We also find qualitative agreement with constraints from abundance matching to numerical simulations \citep{Moster2013} and constraints from a 2$\times$2pt+SMF analysis of COSMOS with a fixed {\it WMAP5} cosmology  \citep[][note here we compare to constraints from the most similar $0.22<z<0.48$ COSMOS sample]{Leauthaud2012} as well as the CFHTLenS/VIPERS analysis by \citet{Coupon2015}. We find our constraints on the scatter of the central CSMF, $\sigma_{\mathrm{c}}$, and the low mass end slope of the satellites, $\alpha_{\mathrm{s}}$, to be in agreement with \citet{Yang2008, Cacciato2012, VanUitert2016, Dvornik2018, Bilicki2021}. The Eddington bias at the high mass end is captured by the $\sigma_{\mathrm{c}}$ parameter, leaving the other parameters mostly unaffected.

We account for the impact of baryon feedback in our model by allowing for freedom in the normalisation of the mass-concentration relation for both the haloes, $f_{\mathrm{h}}$, and the satellite galaxy distribution, $f_{\mathrm{s}}$ (Eq.~\ref{eqn:fsfh}).  With these independent free parameters we can capture the expected small-scale baryon feedback power suppression shown in hydrodynamical simulations \citep{Debackere2021, Amon2022}.   We find $f_{\mathrm{h}}$ and $f_{\mathrm{s}}$ to be consistent with 1, with a preference for lower values, with $1\sigma$ lower limits $f_{\mathrm{h}} > 0.65$ and $f_{\mathrm{s}} > 0.38$.  Our results are consistent with \citet{Viola2015}, indicating that the concentrations of real haloes and satellite distributions are smaller than the haloes in dark matter only simulations \citep[see also][]{Debackere2020, Debackere2021}. Future work will compare direct measurements from hydrodynamical simulations \citep[e.g.][]{McCarthy2017} with our halo model approach to account for the mass dependence of baryonic effects on the radial profiles of dark matter haloes. 

\subsection{Modelling satellite galaxies}
\label{sec:poisson_model}

In our fiducial model we used the findings of \citet{Dvornik2018} that showed that the occupation distribution of satellite galaxies does not follow a Poisson distribution, and that generally the parameter $\mathcal{P}$ (Eq. \ref{beta_def}) is not unity, with our fiducial run preferring a sub-Poissonian behaviour. Following \citet{Cacciato2012}, we  quantify the impact of removing this flexibility in the model, by fixing the parameter $\mathcal{P}$ to unity,
\begin{equation}
\mathcal{P}(M) \equiv  {\langle N_{\mathrm{s}}(N_{\mathrm{s}} - 1) \vert M \rangle \over  \langle N_{\mathrm{s}} \vert M \rangle^{2}} \equiv 1\,,
\end{equation}
thus assuming that the satellite galaxies obey the Poisson distribution. We run another set of MCMC chains using the same setup as in the fiducial case, but with one less parameter to constrain. The resulting constraints are shown in Fig. \ref{fig:full_corner}, with marginalised constraints quoted in Table \ref{tab:results_ns}.  We find significant shifts for the two main cosmological parameters with $\Omega_{\mathrm{m}} = 0.330 {\pm 0.019}$ and $S_{8}=0.951^{+0.037}_{-0.036}$ and a formally acceptable fit with $p(\chi^2 | \nu_{\mathrm{eff}}) = 0.02$ for the whole data vector. In this case we find that the fixed Poisson parameter non-trivially affects the other parameters governing the satellites in the halo model.  Specifically the normalisation of the satellite conditional stellar mass function, $b_0$ and $b_1$, shifts, and these parameters are in turn non-trivially correlated with the main cosmological parameters.  \citet{Cacciato2012} argues that flexibility in the form of the satellite galaxy model is critically important in order to both constrain the galaxy bias \citep{Cacciato2012a, Dvornik2018, Asgari2020a}, and to obtain unbiased cosmological parameters.

\begin{figure}
	\centering
	\includegraphics[width=\columnwidth]{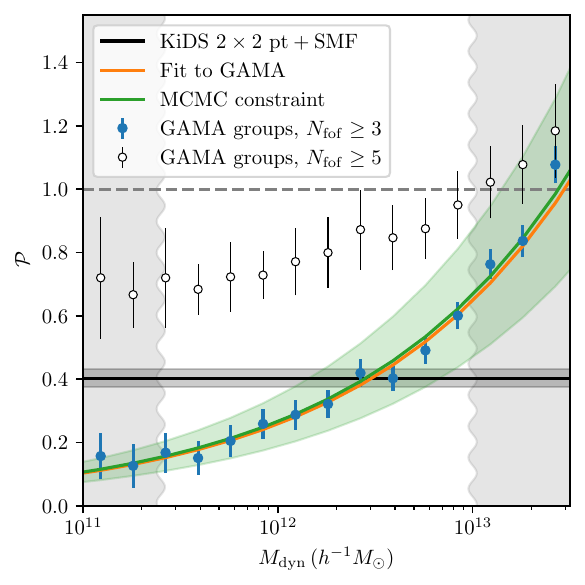}
 	\caption{The Poissonian parameter $\mathcal{P}$ (Eq.~\ref{eqn:PMdyn}), as a function of dynamical group mass, $M_{\mathrm{dyn}}$, for GAMA galaxy groups with 3 or more members (blue), and groups with 5 or more members (white). The orange/green line shows the best fit power law (Eq.~\ref{eqn:PMvar}) to the GAMA/2$\times$2pt+SMF data, with the green region showing the uncertainty on the 2$\times$2pt+SMF fit.
 	 The fiducial, single value, model for  $\mathcal{P}$ is shown with the grey horizontal band. With the wiggly greyed our areas we show the approximate range of dynamical group masses that are outside of halo mass range of our analysis.}
	\label{fig:gama_poisson}
\end{figure}

There are several reasons to reject the results of our Poissonian satellite distribution model analysis. First, whilst we find an acceptable fit of the $\mathcal{P}=1$ model to our full 2$\times$2pt+SMF data vector, there is an unacceptable fit to the $w_{\mathrm{p}}$ part of the data vector, with $p[\chi^2(w_{\mathrm{p}}) | \nu_{\mathrm{eff}}^{w_{\mathrm{p}}}] \sim 10^{-4}$. Secondly, there is observational evidence from the GAMA group catalogue \citep{Robotham2011} that for haloes with masses below $10^{13} h^{-1} M_\odot$, the number of satellites exhibit sub-Poisson behaviour, where in Fig. \ref{fig:gama_poisson} we measure $\mathcal{P}(M)$ as a function of the dynamical mass $M_{\mathrm{dyn}}$ \citep{Driver2022}. We relate the parameter $\mathcal{P}$ with the observed mean and variance of the number of GAMA group members in narrow bins of dynamical mass $M_{\mathrm{dyn}}$ as:
\begin{equation}
    \mathcal{P}(M_{\mathrm{dyn}}) = \left( \frac{\mathrm{Var}[N_{\mathrm{s}}\vert M_{\mathrm{dyn}}]}{\langle N_{\mathrm{s}}\vert M_{\mathrm{dyn}} \rangle} - 1 \right) \frac{1}{\langle N_{\mathrm{s}}\vert M_{\mathrm{dyn}} \rangle} + 1 \,.
    \label{eqn:PMdyn}
\end{equation}
Here the mean $\langle N_{\mathrm{s}}\vert M_{\mathrm{dyn}} \rangle$ and variance $\mathrm{Var}[N_{\mathrm{s}}\vert M_{\mathrm{dyn}}]$ are directly obtained from the GAMA groups catalogue, where we select groups with the number of members that is equal to or larger than 3 \citep{Robotham2011}. We find the satellite distribution to be sub-Poissonian for $M_{\mathrm{dyn}} <10^{13} h^{-1} M_\odot$,  ranging from $\mathcal{P}(M_{\mathrm{dyn}}) = 0.15$ at $M_{\mathrm{dyn}} = 10^{11} h^{-1} M_\odot$, to $\mathcal{P}(M_{\mathrm{dyn}}) = 0.76$ at $M_{\mathrm{dyn}} = 10^{13} h^{-1} M_\odot$, (shown blue in Fig. \ref{fig:gama_poisson}). Assuming the GAMA dynamical mass is a reasonable estimate of the halo mass $M_{\mathrm{h}}$, using Fig.~\ref{fig:smhm} to define our stellar mass range, we expect the satellite distribution to be sub-Poissonian across the full stellar mass range of our 2$\times$2pt+SMF analysis. This sub-Poissonian behaviour is recovered in our fiducial analysis where $\mathcal{P} = 0.403 {\pm 0.029}$. In \citet{Dvornik2018} analysis the behaviour of satellite galaxies is recovered to be super-Poissonian, which is consistent with the trend seen in Fig. \ref{fig:gama_poisson}, as the halo masses of GAMA galaxies in that sample were above $10^{13} h^{-1} M_\odot$.

Finally, hydrodynamical simulations show that for haloes with masses above $\sim 5\times10^{12} h^{-1} M_\odot$, the number of satellites exhibit super-Poisson behaviour \citep{Hadzhiyska2022} and for haloes with masses below $\sim 5\times10^{12} h^{-1} M_\odot$, the number of satellites exhibit sub-Poisson behaviour \citep{Kravtsov2004, Jiang2017, Bhowmick2018, Beltz-Mohrmann2020}. The studies from \citet{Beltz-Mohrmann2020} and \citet{Hadzhiyska2022} show a large uncertainty on the Poisson number and can be to an extent also well described by a model that assumes a Poisson distribution, while the results from \citet{Kravtsov2004}, \citet{Jiang2017} and \citet{Bhowmick2018} show a clear sub-Poisson behaviour for low mass haloes, with the analysis of \citet{Jiang2017} also showing super-Poisson behaviour for high mass haloes with a transition period where satellite galaxies behave completely Poissonian. What is more, a recent analysis \citep{Linke2022} of a semi-analytic simulation \citep{Henriques2015} using galaxy-matter bispectrum shows that parameter $\mathcal{P}$ indeed varies from sub-Poisson behaviour to super-Poisson behaviour as halo mass increases, and it seems to be furthermore dependant on the stellar mass as well (Appendix B therein).

Given the significant impact of the form of the satellite galaxy model on our cosmology constraints, we explore the satellite distribution further, noting that Fig. \ref{fig:gama_poisson} reveals a strong mass dependence that is missing from our fiducial model.  We determine the impact of neglecting this mass-dependence in our analysis, by including an explicit halo mass dependence on the Poisson parameter $\mathcal{P}$. We fit a power law function to the GAMA data as:
\begin{equation}
    \mathcal{P}(M) = A \left(\frac{M}{M_{\mathrm{piv}}} \right)^{B} \,.
    \label{eqn:PMvar}
\end{equation}
finding $A = 0.43$, $B = 0.39$ and $M_{\mathrm{piv}} = 12.54$ when $M=M_{\mathrm{dyn}}$ (shown in orange in Fig. \ref{fig:gama_poisson}). In our extended 2$\times$2pt+SMF analysis, we model $\mathcal{P}(M)$ using Eq.~\ref{eqn:PMvar}, fixing the normalisation $A$ and slope $B$ of the power law to the GAMA best-fit values.  We treat $M_{\mathrm{piv}}$ as a free parameter, however, to account for the uncertainty in the relationship between the GAMA dynamical mass, $M_{\mathrm{dyn}}$, and the true lensing halo mass used in the halo model. The best fit $M_{\mathrm{piv}}$ from the 2$\times$2pt+SMF is nevertheless found to be in good agreement with the GAMA fit (shown in green in Fig. \ref{fig:gama_poisson}).

Using the $\mathcal{P}(M)$ model we obtain the parameter constraints listed in Table \ref{tab:results} and shown in Fig \ref{fig:full_corner}. We find a good fit of the model to the data with $p(\chi^2 | \nu_{\mathrm{eff}}) = 0.19$.
All the parameters are consistent with the constraints from our fiducial analysis that assumes no mass dependence.  We find a preference for lower values for the two primary cosmological parameters, with $S_8=0.718^{+0.045}_{-0.040}$  and $\Omega_{\mathrm{m}} = 0.276^{+0.024}_{-0.021}$ corresponding to a $1.3\sigma$ and $0.7\sigma$ difference from the fiducial result (see also Fig. \ref{fig:s8_summary_tests}).

This extension analysis shows that our results are sensitive to how we model the mass-dependence of the satellite group distribution, at an acceptable level of $\sim 1\sigma$ in our primary cosmological parameters.  We chose not to use this extended $\mathcal{P}(M)$ model in our fiducial analysis, as in Fig. \ref{fig:gama_poisson}, we show how sensitive the GAMA-measured $\mathcal{P}(M_{\mathrm{dyn}})$ relationship is to the group selection criteria.  We find that the behaviour changes when groups are defined with a number of members that is equal or larger than 5 (white data points, compared to the blue data points for our original selection criteria).  In future higher signal-to-noise studies we will explore keeping the parameters $A$ and $B$ free in the $\mathcal{P}(M)$ model, and implement a more complex model that uses the non-Poisson behaviour directly in the definition of the HOD \citep[for instance a negative binomial distribution as shown in][]{Boylan-Kolchin2010}. Further considerations need to be taken into account for possible stellar mass dependency of the Poisson parameter $\mathcal{P}$ \citep{Linke2022}.

\section{Discussion and conclusions}
\label{sec:conclusions}

In this paper we combined measurements of galaxy clustering, galaxy-galaxy lensing and galaxy abundances in the form of the stellar mass function, in order to simultaneously set constraints on cosmological parameters and galaxy bias. Using a flexible halo model, we analysed the fourth data release of the Kilo-Degree Survey (KiDS) \citep[KiDS-1000,][]{Kuijken2019}, where the source sample, used to measure the galaxy-galaxy lensing signal, has undergone a rigorous study to assess robustness and accuracy \citep{Asgari2020, Giblin2020, Hildebrandt2020}.  For our lens sample we used the KiDS-Bright sample \citep{Bilicki2021} whose selection was calibrated against a complete and representative spectroscopic sample from the GAMA survey \citep{Driver2011}, with the photometric redshifts calibrated using a neural network \citep[{\sc ANNz2},][]{Sadeh2016}. The resulting accuracy of the estimated redshifts for the KiDS-bright sample is sufficient for galaxy-galaxy lensing and galaxy clustering studies.

We used the halo model to analyse our data, building upon the cosmological analyses presented in \citet{Cacciato2012} and \citet{More2014}. We used a single halo occupation model to compute the clustering of galaxies and the galaxy-galaxy lensing signal \citep{Guzik2002, Yoo2006, Cacciato2009}, and the galaxy abundances \citep{Bosch2012, Cacciato2012}. This model was shown to be able to simultaneously constrain cosmology and halo occupation statistics, as well as constrain the extensions to the standard $\Lambda$CDM cosmologies, such as the equation of state of dark energy and neutrino mass \citep{More2012, Krause2016}.

We improved upon previous studies by (i) using a more accurate N-body simulation calibrated analytical model, taking into account halo exclusion, scale dependence and the non-linear nature of halo bias \citep{Mead2021, Mahony2022}, (ii) using the latest lensing and clustering data from a single survey (KiDS-1000), and (iii) using the full analytical covariance matrix that accounts for cross-covariance between all observables and in particular the cross-covariance between the stellar mass function and two point statistics. 

We have adopted a Bayesian approach to constrain our model parameters, using the MCMC to probe the posterior distributions. For a flat $\Lambda$CDM cosmology we find $\Omega_{\mathrm{m}} = 0.290^{+0.021}_{-0.017}$ and $S_8 = 0.773^{+0.028}_{-0.030}$, which is consistent and comparable to constraints from 3$\times$2pt studies that also include a cosmic shear observable \citep{Heymans2020, DESY3}.  Our results follow the trend seen in other lensing studies for a tension in $S_8$ when compared to \citet{PlanckCollaboration2018}.  Using the Hellinger distance as a tension metric, this difference is at the 1.9$\sigma$ level for our 2$\times$2pt+SMF analysis. We find that our constraints are sensitive, at the $\sim 1\sigma$-level (in $S_8$), to how we choose to model the mass-dependence of the satellite distribution within the halo model.  This aspect of our analysis will require further development in future higher signal-to-noise studies.

Combining galaxy clustering and galaxy-galaxy lensing with cosmic shear measurements has been a standard approach for large-scale structure analyses in recent years \citep{VanUitert2018, Joudaki2017, DESY1,DESY3, Heymans2020}. We anticipate that combing our halo model approach with cosmic shear data will allow for additional constraints on astrophysical systematics arising from the intrinsic alignment of galaxies and baryon feedback. So far, intrinsic alignments in cosmic shear analysis have either been included using a non-linear modification of the linear alignment model \citep[NLA]{Bridle2007} or a perturbation theory approach \citep[TATT][]{Blazek2019}. For a consistent halo model approach this effect could also be modelled within the framework adopted in this analysis \citep[e.g.][]{Fortuna2020}. In this analysis we have varied the halo concentration parameter to account for baryon feedback.  With additional data, a more complex halo model could be adopted allowing for the inclusion of gas observables to constrain baryon feedback through the Sunyaev-Zeldovich effect \citep{MeadHmX,Troester2022}.  The flexibility of the halo model also allows for extensions to the underlying cosmological model without having to employ a myriad of costly simulations to cover a large range of parameters \citep{Cataneo2019}.  We therefore see a significant role for our adopted methodology in future cosmological analyses of upcoming large-scale structure surveys.

\begin{acknowledgements}
We thank Andrew Hearin for useful discussion on modelling the non-Poissonian behaviour of satellite galaxies, and the anonymous referee for the helpful comments and constructive remarks on this manuscript. We also thank Constance Mahony for spotting the issue with clustering measurements presented in the original published paper. We acknowledge support from the European Research Council under grants 770935 (AD, CM, HHi, RR, AHW) and 647112 (CH, MA), and the UK Science and Technology Facilities Council (STFC) under grants ST/V000594/1 (CH, MA) and ST/V000780/1 (BJ). HHi is further supported by a Heisenberg grant of the Deutsche Forschungsgemeinschaft (Hi 1495/5-1). CH and AM acknowledge support from the Max Planck Society and the Alexander von Humboldt Foundation in the framework of the Max Planck-Humboldt Research Award endowed by the Federal Ministry of Education and Research. MB is supported by the Polish National Science Center through grants no. 2020/38/E/ST9/00395, 2018/30/E/ST9/00698, 2018/31/G/ST9/03388 and 2020/39/B/ST9/03494, and by the Polish Ministry of Science and Higher Education through grant DIR/WK/2018/12. EC and HJ acknowledges support from the Delta ITP consortium, a program of the Netherlands Organisation for Scientific Research (NWO) that is funded by the Dutch Ministry of Education, Culture and Science (OCW), project number 24.001.027. HHo acknowledges support from Vici grant 639.043.512, financed by the Netherlands Organisation for Scientific Research (NWO). KK acknowledges support from the Royal Society and Imperial College. HM was supported in part by World Premier International Research Center Initiative (WPI Initiative), MEXT, Japan, by JSPS KAKENHI Grant Numbers 20H01932, by JSPS Core-to-Core Program Grant Number JPJSCCA20200002, and by Japan Science and Technology Agency (JST) CREST JPMHCR1414 and JST AIP Acceleration Research Grant Number JP20317829, Japan. TN is supported in part by MEXT/JSPS KAKENHI Grant Numbers P19H00677, P20H05861, JP21H01081, JP22K0363, and Japan Science and Technology Agency (JST) AIP Acceleration Research Grant Number JP20317829. \\

Based on observations made with ESO Telescopes at the La Silla Paranal Observatory under programme IDs 177.A-3016, 177.A-3017, 177.A-3018 and 179.A-2004, and on data products produced by the KiDS consortium. The KiDS production team acknowledges support from: Deutsche Forschungsgemeinschaft, ERC, NOVA and NWO-M grants; Target; the University of Padova, and the University Federico II (Naples).\\

This work has made use of Python (\url{http://www.python.org}), including the packages \texttt{numpy} (\url{http://www.numpy.org}), \texttt{scipy} (\url{http://www.scipy.org}), \texttt{astropy} \citep[\url{http://www.astropy.org}][]{Astropy2013, Astropy2018}, and \texttt{hmf} \citep{Murray2013}. Plots have been produced with \texttt{matplotlib} \citep{Hunter2007} and \texttt{chainconsumer} \citep{Hinton2016}. This work has made use of CosmoHub for validation of our covariance matrices. CosmoHub has been developed by the Port d'Informació Científica (PIC), maintained through a collaboration of the Institut de Física d'Altes Energies (IFAE) and the Centro de Investigaciones Energéticas, Medioambientales y Tecnológicas (CIEMAT) and the Institute of Space Sciences (CSIC \& IEEC), and was partially funded by the “Plan Estatal de Investigación Científica y Técnica y de Innovación” program of the Spanish government. \\
\\

\\
\textit{Author contributions:} All authors contributed to writing and development of this paper. The authorship list reflects the lead authors (AD,CH,MA,CM,BJ) followed by an alphabetical group that includes those who are key contributors to the scientific analysis. For the purpose of open access, the author has applied a Creative Commons Attribution (CC BY) licence to any Author Accepted Manuscript version arising from this submission.

\end{acknowledgements}

\bibliographystyle{aa}
\bibliography{library,extra_refs}

\begin{appendix}
\onecolumn

\section{Stellar mass function covariance matrix}
\label{sec:smf_cov}

We derive the covariance of the SMF in direct analogy to the flux-limited case considered in \citet{Smith2012}, but neglect the halo occupation variance contribution because it was demonstrated to be always subdominant \citep{Smith2012}. To simplify the expression we neglect the mapping from true stellar mass to observed stellar mass, where the corresponding integrations would appear explicitly as this relation is expected to be fairly tight. The SMF covariance is composed of a shot noise and super-sample covariance (SSC) contribution,
\begin{equation}
\mathrm{Cov}\left[\Phi_{\mu}^{i}, \Phi_{\nu}^{j}\right] = \mathrm{Cov}^{\mathrm{SN}}\left[\Phi_{\mu}^{i}, \Phi_{\nu}^{j}\right] + \mathrm{Cov}^{\mathrm{SSC}}\left[\Phi_{\mu}^{i}, \Phi_{\nu}^{j}\right]\,,
\end{equation}
with
\begin{equation}
\mathrm{Cov}^{\mathrm{SN}}\left[\Phi_{\mu}^{i}, \Phi_{\nu}^{j}\right] = \delta_{i,j}\,\delta_{\mu,\nu} \frac{\Phi_{\mu}^{i}}{\Delta M_{\star} V_{\mathrm{max},\mu}^{i}}\,,
\end{equation}
where $\Phi_{\mu}^{i} = \Phi^{i}(M_{\star, \mu})$ and $V_{\mathrm{max},\mu}^{i} = V_{\mathrm{max}}^{i}(M_{\star, \mu})$ are the shorthands for the stellar mass function and $V_{\mathrm{max}}$ of stellar mass bin $\mu$ and generally a tomographic bin $i$. We define:
\begin{equation}
\widetilde{\Phi}_{\mu} = \int_{0}^{\infty} \Phi(M_{\star, \mu} \vert M, [z]) \, n(M, z)\,b_{\mathrm{h}}(M,z) \, \mathrm{d} M\,,
\end{equation}
using which the SSC terms are given by
\begin{equation}
\mathrm{Cov}^{\mathrm{SSC}}\left[\Phi_{\mu}^{i}, \Phi_{\nu}^{j}\right] = \frac{A_{\mathrm{survey}}^{2}\, f^{i} f^{j}}{V_{\mathrm{max},\mu}^{i}\, V_{\mathrm{max},\nu}^{j}} \int \mathrm{d}\chi \frac{p^{i}(\chi)}{p_{\mathrm{tot}}(\chi)} \frac{p^{j}(\chi)}{p_{\mathrm{tot}}(\chi)}\, f^{2}_{K}(\chi)\, \sigma^{2}_{\mathrm{s}}(\chi)\,\widetilde{\Phi}_{\mu}[z(\chi)]\,\widetilde{\Phi}_{\mu}[z(\chi)] \,,
\end{equation}
where $\sigma^{2}_{\mathrm{s}}$ is the variance of density fluctuations within the angular survey window (see Appendix E of \citet{Joachimi2020} for definition), $p^{i}$ are the tomographic bin-wise redshift distributions, $p_{\mathrm{tot}}$ the overall redshift distribution for all galaxies in the sample/survey, $f_{K}$ the comoving angular diameter distance, $\chi$ the comoving radial distance and $f^{i}$ the fraction of galaxies in bin $i$ relative to all galaxies. $A_{\mathrm{survey}}$ is the survey area.

The cross-variance is derived in close analogy to \citet{Takada2007}, with the consistency checks by \citet{Schaan2014}. The cross-variance receives contributions from two terms,
\begin{equation}
\mathrm{Cov}\left[\Phi_{\mu}^{i}, \mathcal{O}^{jl}(r_{\mathrm{p}})\right] = \mathrm{Cov}^{\mathrm{CV}}\left[\Phi_{\mu}^{i}, \mathcal{O}^{jl}(r_{\mathrm{p}})\right] + \mathrm{Cov}^{\mathrm{SSC}}\left[\Phi_{\mu}^{i}, \mathcal{O}^{jl}(r_{\mathrm{p}})\right]\,,
\end{equation}
a cosmic variance (CV) and a super-sample covariance (SSC) term, respectively. Here we determine the cross-variance with a projected two-point function $\mathcal{O}^{jl}(r_{\mathrm{p}})$ either WP, $w_{\mathrm{p}}(r)$, or ESD, $\Delta\Sigma(r)$, in bins $j$ and $l$, respectively. The cosmic variance contribution is a three-point correlation given by
\begin{equation}
\mathrm{Cov}^{\mathrm{CV}}\left[\Phi_{\mu}^{i}, \mathcal{O}^{jl}(r_{\mathrm{p}})\right] = \overline{\rho}_{\mathrm{x}} f^{i} \int \mathrm{d}\chi \frac{p^{i}(\chi)}{p_{\mathrm{tot}}(\chi)} \int {\text{d}k\,k \over 2 \pi} J_{\mathrm{x}}(kr_{\text{p}})\, \left(nB_{\mathrm{cmm}}\right)_{\mu}^{jl} \left[k, z(\chi)\right] \,,
\end{equation}
where $\overline{\rho}_{\mathrm{x}} = 1$ and $J_{\mathrm{x}} = J_{\mathrm{0}}$ in the case when observable is $w_{\mathrm{p}}$, and $\overline{\rho}_{\mathrm{x}} = \overline{\rho}_{\mathrm{m}}$ and $J_{\mathrm{x}} = J_{\mathrm{2}}$ in the case when observable is $\Delta\Sigma$. $J_{n}$ are Bessel functions of $n$-th kind. Here we have defined the count-matter cross-bispectrum (evaluated for a collapsed triangle) in close analogy to \citet{Takada2007}. It can be expressed in the halo model formalism as
\begin{align}
\left(nB_{\mathrm{cmm}}\right)_{\mu} \left(k, z\right) &= \int \mathrm{d}M \, n(M,z) \, \Phi(M_{\star, \mu} \vert M, [z]) \, \left({M \over \overline{\rho}_{\mathrm{m}}}\right)^{2} \, \tilde{u}^{2}_{\mathrm{h}}(k \vert M) \\ \nonumber
&+ 2P_{\mathrm{lin}}(k,z) \int \mathrm{d}M \, n(M,z) \, \Phi(M_{\star, \mu} \vert M, [z]) \, b_{\mathrm{h}}(M,z) {M \over \overline{\rho}_{\mathrm{m}}} \, \tilde{u}_{\mathrm{h}}(k \vert M) \int \mathrm{d}M^{\prime} \, n(M^{\prime},z) \, b_{\mathrm{h}}(M^{\prime},z) {M^{\prime} \over \overline{\rho}_{\mathrm{m}}} \, \tilde{u}_{\mathrm{h}}(k \vert M^{\prime})\,.
\end{align}
Finally, the SSC term reads
\begin{equation}
\mathrm{Cov}^{\mathrm{SSC}}\left[\Phi_{\mu}^{i}, \mathcal{O}^{jl}(r_{\mathrm{p}})\right] = \overline{\rho}_{\mathrm{x}} A_{\mathrm{survey}} f^{i} \int \mathrm{d}\chi \frac{p^{i}(\chi)}{p_{\mathrm{tot}}(\chi)} \int {\text{d}k\,k \over 2 \pi} J_{\mathrm{x}}(kr_{\text{p}})\, \frac{\partial P_{\mathrm{xy}}^{jl}(k, z(\chi))}{\partial \delta_{\mathrm{b}}}\, \sigma^{2}_{\mathrm{s}}(\chi)\, \widetilde{\Phi}_{\mu}[z(\chi)]\,,
\end{equation}
where $P_{\mathrm{xy}}(k, z)$ is either $P_{\mathrm{gg}}(k, z)$ for $w_{\mathrm{p}}$ or $P_{\mathrm{gm}}(k, z)$ for $\Delta\Sigma$, and the derivative is with respect to a super-survey density fluctuation $\delta_{\mathrm{b}}$ \citep{Takada2013, Dvornik2018}. As the systematic and statistical uncertainties on stellar masses are comparable in power \citep{Brouwer2021}, the entries in the SMF and cross-covariances are inflated by a factor of 2 to account for uncertainty arising from Eddington bias and the systematic shift in stellar masses.

\twocolumn

\section{Prior space}
\label{sec:prior}

\begin{figure}
	\centering
	\includegraphics[width=\columnwidth]{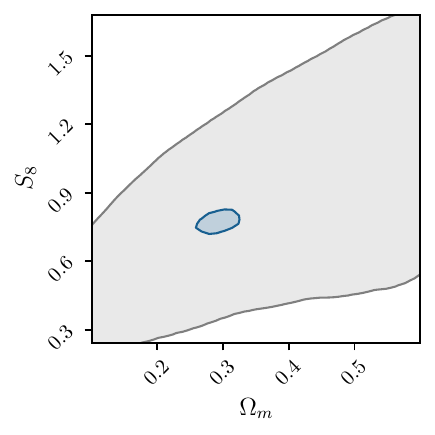}
 	\caption{Prior range for the $\Omega_{\mathrm{m}}$ and $S_8$ parameters compared to the $2\sigma$ contour from our fiducial cosmological analysis.}
	\label{fig:prior_range}
\end{figure}

We adopt informative priors for three cosmological parameters; $n_{\mathrm{s}}$, $\Omega_{\mathrm{b}}$, and $h$.  Our priors are also informative on the parameters which scale the mass-concentration relation for haloes and satellite galaxies; $f_{\mathrm{h}}$ and $f_{\mathrm{s}}$.  In Fig. \ref{fig:prior_range} we verify that our choice of priors does not inform the $\Omega_{\mathrm{m}}$ and $S_8$ parameters, noting that our prior combination does not result in trivial square prior coverage.

\section{Estimating the goodness of fit}
\label{sec:fit}

To assess the goodness-of-fit of our model to the data it is necessary to determine the effective number of degrees of freedom, $\nu_{\mathrm{eff}}$, \citep[see the discussion in section 6.3 of][on why $\nu_{\mathrm{eff}} \neq N_{\mathrm{data}} -  {\mathrm{N}}_{\theta}$, for a typical cosmology analysis with ${\mathrm{N}}_{\mathrm{data}}$ data points and $ {\mathrm{N}}_{\theta}$ cosmological parameters]{Joachimi2020}.  We follow a simulation approach to determine $\nu_{\mathrm{eff}}$ following \citet{Joachimi2020, Miyatake2021}. 

\begin{figure}
	\centering
	\includegraphics[width=\columnwidth]{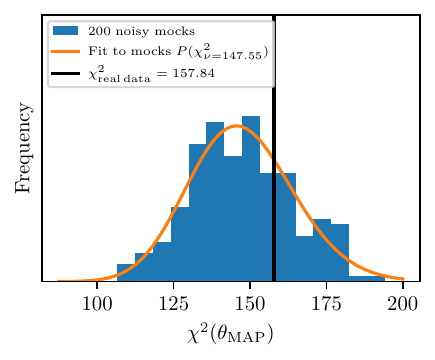}
 	\caption{An estimation of goodness of fit of the fiducial best fit model at the maximum a posteriori (MAP) values. The histogram shows the distribution of the $\chi^2$ values from 200 noisy mock data vectors (see text for detailed procedure). The orange line shows the fit of the $\chi^2$ distribution to the histogram, obtaining the effective number of degrees of freedom in the data. The black vertical line shows the $\chi^2$ value as obtained from the best fit model to the real data.}
	\label{fig:chi}
\end{figure}

We generate 200 noisy mock data vectors, drawing samples from the multivariate distribution defined by the mean, which is our best-fit signal, and the full covariance matrix. We find the maximum posterior point using a Nelder-Mead minimisation and its corresponding $\chi^2$ value for the fit. From the resulting distribution of $\chi^2$ values, shown in Fig.~\ref{fig:chi}, we fit a $\chi^2$ distribution to find the effective degrees of freedom\footnote{We note that our simulation approach finds a slightly larger $\nu_{\mathrm{eff}}$ compared to the estimation using the \citet{Raveri2019} approach, for which we find $\nu_{\mathrm{eff}}=138.5$,}, $\nu_{\mathrm{eff}}=147.55$, finding that our model provides a good fit to the data with $p(\chi^2 | \nu_{\mathrm{eff}})=0.27$.

\begin{figure}
	\centering
	\includegraphics[width=\columnwidth]{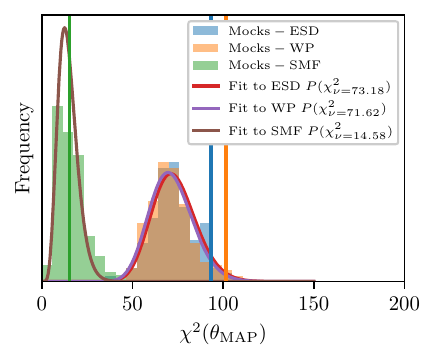}
 	\caption{Same as Fig. \ref{fig:chi}, but for the 3 different observables in our analysis. The fits to the distributions are used to determine the number of degrees of freedom for each observable which are in turn are used to determine the reduced $\chi^2$ values for each of them respectively. Note that the $\chi^2$ distributions are not a result of a maximising the posterior for that section of the mocks. The vertical lines show the $\chi^2$ values from the real data, matching in colours with the histograms.}
	\label{fig:chi_parts}
\end{figure}

We can use our simulation approach to also determine the goodness of fit for each of the three sections of the data vector: ESD (galaxy-galaxy lensing), $w_{\mathrm{p}}$ (galaxy clustering), and SMF (stellar mass function). To estimate the degrees of freedom for each observable we find the $\chi^2$ value for that section of the data vector corresponding to the maximum posterior point that is found using the full data vector. In other words the partial $\chi^2$ values are not separately minimised.  Fig.~\ref{fig:chi_parts} presents the resulting $\chi^2$ distributions and fit.  We note that some alternative methods to define $\nu_{\mathrm{eff}}$, such as \citet{Raveri2019}, would be ill defined for these sub-data vectors owing to the number of free parameters in the model, resulting in negative effective degree of freedom or negative effective number of parameters. We find $\nu^{\mathrm{ESD}}_{\mathrm{eff}}=73.18$ with $p[\chi^2 ({\mathrm{ESD}}) | \nu^{\mathrm{ESD}}_{\mathrm{eff}}]=0.05$, $\nu^{w_{\mathrm{p}}}_{\mathrm{eff}}=72.62$ with $p[\chi^2 ({w_{\mathrm{p}}}) | \nu^{w_{\mathrm{p}}}_{\mathrm{eff}}]=0.01$, and $\nu^{\mathrm{SMF}}_{\mathrm{eff}}=14.58$ with $p[\chi^2 ({\mathrm{SMF}}) | \nu^{\mathrm{SMF}}_{\mathrm{eff}}]=0.39$.  We conclude that the goodness of fit of the full data vector and each of the individual sections is acceptable.  This procedure is repeated for the remaining modelling cases considered in this paper.

\section{Galaxy-galaxy lensing random signal}
\label{sec:randoms}

\begin{figure}
	\centering
	\includegraphics[width=\columnwidth]{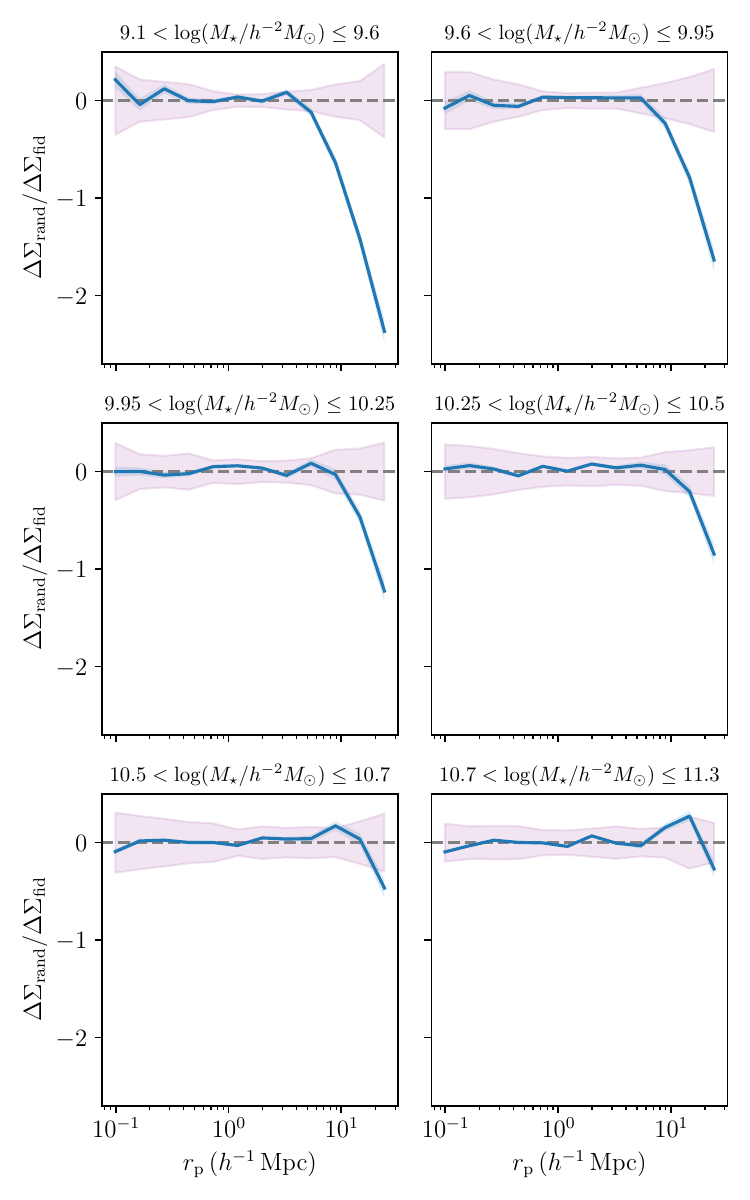}
 	\caption{The ratio between the lensing signal measured around random galaxies from the organised randoms catalogue and the fiducial model prediction in our six stellar mass bins. For comparison we show the uncertainty of the data as a purple band.}
	\label{fig:random}
\end{figure} 

The ratio between the lensing signal measured around random galaxies from the organised randoms catalogue and the fiducial model prediction in our six stellar mass bins is presented in Fig. \ref{fig:random}. A non-zero random signal indicates systematic effects in the ESD signal measured from the KiDS survey. The strong dependence of the random signal on the stellar mass of the lens bin indicates that this systematic is unlikely to be caused by residual uncorrected distortions in the source lensing catalogue, for example those associated with the point spread function of the camera.  Such a data-related systematic would impact the ESD signal of each lens bin similarly given the similar source samples \citep[see also][which presents a series of diagnostic tests]{Giblin2020}.  Instead we conclude that this signal arises from large scale structure sample variance that differs as the volume of the sample grows with the increasing stellar mass \citep{Singh2016}.  We calculate the error on the mean random signal, measured using 1000 samples of the random catalogue, finding it to be indistinguishable from the thickness of the blue curve, and sufficiently small that we do not include it in our overall error budget.  We find that the random correction applied to the data exceeds 100\% in the first three stellar mass bins on large scales, and can be up to 4 times larger than the statistical uncertainty.

\section{Extensions to the fiducial cosmological analysis}
\label{sec:tests}

\begin{figure}
	\centering
	\includegraphics[width=\columnwidth]{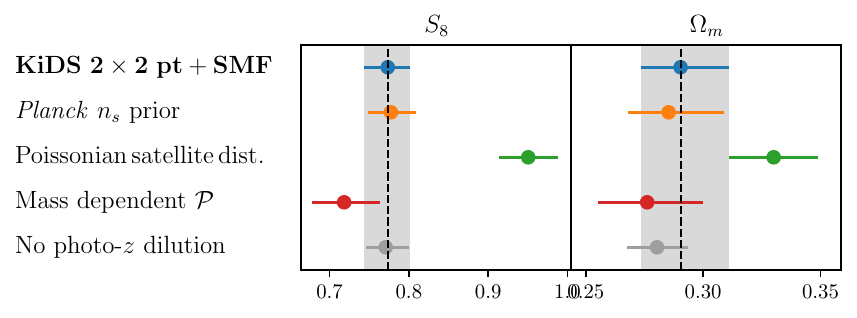}
 	\caption{Comparison between $S_8$ and $\Omega_{\mathrm{m}}$ values for our fiducial results and the tests for different modelling choices. All results are shown for maximum statistics of the marginal posterior distributions (MMAX) and corresponding credible interval.}
	\label{fig:s8_summary_tests}
\end{figure}

\begin{figure*}
	\centering
	\includegraphics[width=\textwidth]{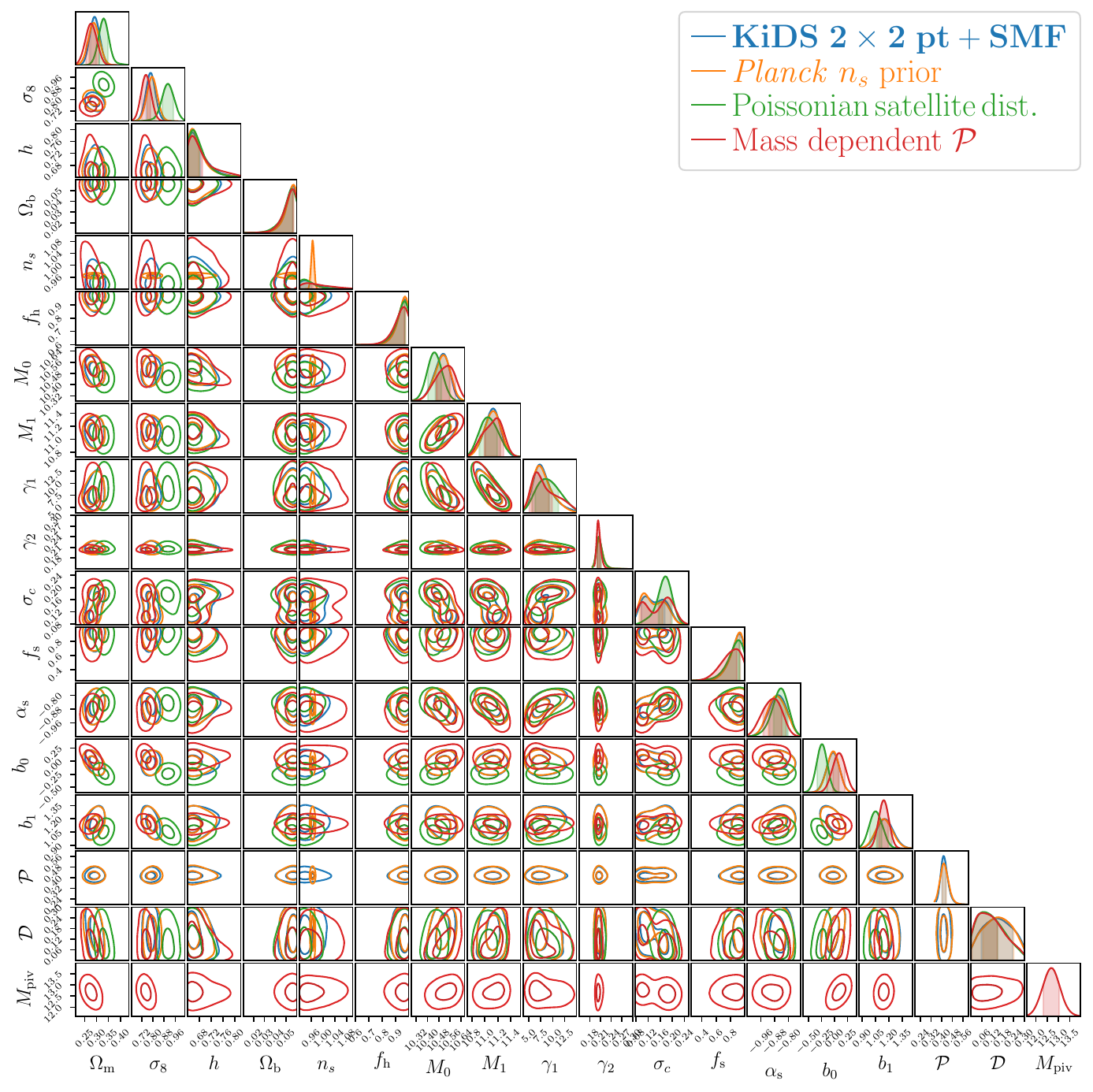}
 	\caption{The full posterior distributions of the model parameters (where the priors are listed in Table \ref{tab:results}). The contours indicate $1\sigma$ and $2\sigma$ confidence regions.}
	\label{fig:full_corner}
\end{figure*}

In this appendix we review the impact of modelling choices on our fiducial cosmological parameter constraints (Sect.~\ref{sec:results}) to a series of extensions to our fiducial theoretical model.  The results, in terms of $S_8$ and $\Omega_{\mathrm{m}}$ are summarised in Fig.~\ref{fig:s8_summary_tests} with the full parameter space quantified in Table~\ref{tab:results_ns} and Fig.~\ref{fig:full_corner}.

\begin{table*}
    \fontsize{8.5pt}{10.25pt}\selectfont
    \centering 
    \caption{Marginal constraints on all model parameters listed together with their priors, for fiducial and extension setups considered.}
    \label{tab:results_ns}
    \setlength\tabcolsep{4pt}
    \begin{tabular}{cccccccccc}
        \toprule
        Parameter & Prior & \multicolumn{2}{c}{Fiducial} & \multicolumn{2}{c}{\textit{Planck} $n_{s}$ prior} & \multicolumn{2}{c}{Poissonian satellite dist.} & \multicolumn{2}{c}{Mass dependent $\mathcal{P}$}  \\ 
        & & MMAX & MAP+PJ-HPD & MMAX & MAP+PJ-HPD & MMAX & MAP+PJ-HPD & MMAX & MAP+PJ-HPD \\
        \midrule
        $\Omega_{\mathrm{m}}$ & $[0.1, 0.6]$ & $0.290^{+0.021}_{-0.017}$ & $0.307^{+0.002}_{-0.031}$ & $0.285^{+0.024}_{-0.017}$ & $0.289^{+0.010}_{-0.026}$ & $0.330\pm 0.019$ & $0.323^{+0.019}_{-0.011}$ & $0.276^{+0.024}_{-0.021}$ & $0.291^{+0.004}_{-0.038}$ \\  \addlinespace
        $\sigma_{8}$ & $[0.4, 1.2]$ & $0.781^{+0.033}_{-0.029}$ & $0.801^{+0.013}_{-0.041}$ & $0.792^{+0.035}_{-0.031}$ & $0.791^{+0.038}_{-0.020}$ & $0.902^{+0.044}_{-0.034}$ & $0.953^{+0.009}_{-0.067}$ & $0.750^{+0.033}_{-0.031}$ & $0.731^{+0.054}_{-0.002}$ \\ \addlinespace
        $h$ & $[0.64, 0.82]$ & $<0.726$ & $<0.711$ & $<0.716$ & $<0.715$ & $<0.720$ & $<0.712$ & $<0.720$ & $<0.720$ \\ \addlinespace
        $\Omega_{\mathrm{b}}$ & $[0.01, 0.06]$ & $>0.01$ & $>0.01$ & $>0.01$ & $>0.01$ & $>0.01$ & $>0.01$ & $>0.01$ & $>0.01$ \\ \addlinespace
        $n_{s}$ & $[0.92, 1.1]$ & $<1.004$ & $<0.978$ & $0.964^{+0.005}_{-0.004}$ & $0.965^{+0.003}_{-0.005}$ & $<0.988$ & $<0.979$ & $<0.996$ & $<0.998$ \\
        \midrule
        $S_8$ & -- & $0.773^{+0.028}_{-0.030}$ & $0.809^{+0.001}_{-0.055}$ & $0.777^{+0.031}_{-0.029}$ & $0.776^{+0.030}_{-0.023}$ & $0.951^{+0.037}_{-0.036}$ & $0.989^{+0.020}_{-0.057}$ & $0.718^{+0.045}_{-0.040}$ & $0.720^{+0.048}_{-0.030}$ \\
        \midrule
        $f_{\mathrm{h}}$ & $[0.0, 1.0]$ & $>0.645$ & $>0.939$ & $>0.665$ & $>0.948$ & $>0.652$ & $>0.938$ & $>0.581$ & $>0.928$ \\ \addlinespace
        $M_0$ & $[7.0, 13.0]$ & $10.519^{+0.039}_{-0.062}$ & $10.521^{+0.062}_{-0.044}$ & $10.520^{+0.042}_{-0.060}$ & $10.537^{+0.051}_{-0.054}$ & $10.446^{+0.053}_{-0.044}$ & $10.511^{+0.010}_{-0.088}$ & $10.556^{+0.044}_{-0.090}$ & $10.595^{+0.051}_{-0.102}$ \\ \addlinespace
        $M_1$ & $[9.0, 14.0]$ & $11.138^{+0.099}_{-0.132}$ & $11.145^{+0.136}_{-0.098}$ & $11.129^{+0.115}_{-0.126}$ & $11.168^{+0.080}_{-0.141}$ & $11.039^{+0.153}_{-0.123}$ & $11.197^{+0.033}_{-0.238}$ & $11.205^{+0.083}_{-0.211}$ & $11.240^{+0.120}_{-0.193}$ \\ \addlinespace
        $\gamma_1$ & $[2.5, 15.0]$ & $7.096^{+2.144}_{-1.406}$ & $7.385^{+1.441}_{-1.824}$ & $7.026^{+2.157}_{-1.454}$ & $6.774^{+2.174}_{-1.014}$ & $8.364^{+2.831}_{-2.028}$ & $6.334^{+4.370}_{-0.343}$ & $6.565^{+3.298}_{-1.494}$ & $6.148^{+3.397}_{-1.104}$\\ \addlinespace
        $\gamma_2$ & $[0.0, 10.0]$ & $0.201 {\pm 0.010}$ & $0.201 {\pm 0.009}$ & $0.201 {\pm 0.011}$ & $0.211 {\pm 0.010}$ & $0.201 {\pm 0.010}$ & $0.207 {\pm 0.007}$ & $0.201 {\pm 0.007}$ & $0.203 {\pm 0.004}$\\ \addlinespace
        $\sigma_{\mathrm{c}}$ & $[0.0, 2.0]$ & $0.108^{+0.067}_{-0.011}$ & $0.159^{+0.007}_{-0.070}$ & $0.105^{+0.062}_{-0.015}$ & $0.098^{+0.063}_{-0.014}$ & $0.178^{+0.021}_{-0.023}$ & $0.123^{+0.062}_{-0.008}$ & $0.184^{+0.011}_{-0.088}$ & $0.095^{+0.087}_{-0.014}$ \\ \addlinespace
        $f_{\mathrm{s}}$ & $[0.0, 1.0]$ & $>0.377$ & $>0.84$ & $>0.394$ & $>0.858$ & $>0.223$ & $>0.808$ & $>0.288$ & $>0.757$\\ \addlinespace
        $\alpha_{\mathrm{s}}$ & $[-5.0, 5.0]$ & $-0.858^{+0.048}_{-0.052}$ & $-0.847^{+0.013}_{-0.097}$ & $-0.874^{+0.059}_{-0.042}$ & $-0.898^{+0.062}_{-0.043}$ & $-0.844^{+0.042}_{-0.043}$ & $-0.916^{+0.092}_{-0.002}$ & $-0.879^{+0.041}_{-0.067}$ & $-0.962^{+0.096}_{-0.035}$ \\ \addlinespace
        $b_0$ & $[-5.0, 5.0]$ & $-0.024^{+0.108}_{-0.117}$ & $-0.120^{+0.199}_{-0.001}$ & $-0.016^{+0.098}_{-0.118}$ & $-0.077^{+0.163}_{-0.028}$ & $-0.239 {\pm 0.092}$ & $-0.267^{+0.104}_{-0.058}$ & $0.093^{+0.107}_{-0.121}$ & $0.029^{+0.141}_{-0.069}$ \\ \addlinespace
        $b_1$ & $[-5.0, 5.0]$ & $1.149^{+0.091}_{-0.081}$ & $1.177^{+0.058}_{-0.096}$ & $1.139^{+0.091}_{-0.078}$ & $1.181^{+0.037}_{-0.111}$ & $1.045^{+0.076}_{-0.060}$ & $1.019^{+0.092}_{-0.029}$ & $1.147^{+0.048}_{-0.057}$ & $1.152^{+0.026}_{-0.073}$ \\ \addlinespace
        $\mathcal{P}$ & $[0.0, 2.0]$ & $0.403 {\pm 0.029}$ & $0.417^{+0.024}_{-0.013}$ & $0.403 {\pm 0.030}$ & $0.411^{+0.014}_{-0.011}$ & 1.0 fixed & 1.0 fixed & -- & -- \\ \addlinespace
        $M_{\mathrm{piv}}$ & $[8.0, 15.0]$ & -- & -- & -- & -- & -- & -- & $12.681^{+0.351}_{-0.361}$ & $12.512^{+0.383}_{-0.290}$ \\ \addlinespace
        $\mathcal{D}$ & $[0.0, 0.3]$ & $0.144^{+0.091}_{-0.085}$ & $0.051^{+0.172}_{-0.006}$ & $0.150^{+0.090}_{-0.082}$ & $0.195^{+0.029}_{-0.147}$ & $0.067^{+0.082}_{-0.064}$ & $0.140^{+0.016}_{-0.132}$ & $0.103^{+0.145}_{-0.073}$ & $0.100^{+0.119}_{-0.058}$ \\
        \midrule
        $\chi^2_{\mathrm{red}}$ & -- & \multicolumn{2}{c}{1.07} & \multicolumn{2}{c}{1.07} & \multicolumn{2}{c}{1.25} & \multicolumn{2}{c}{1.1} \\ \addlinespace
        $p$-value & -- & \multicolumn{2}{c}{0.27} & \multicolumn{2}{c}{0.27} & \multicolumn{2}{c}{0.02} & \multicolumn{2}{c}{0.19} \\ \addlinespace

        \bottomrule
        \addlinespace
    \end{tabular}
    
    \caption*{Notes: This table lists all the free parameters in our model:  the energy density of cold matter $\Omega_{\mathrm{m}}$, the normalisation of power spectrum  $\sigma_8$, the dimensionless Hubble parameter $h$, the spectral index $n_s$, the energy density of baryonic matter $\Omega_{\mathrm{b}}$, the derived parameter $S_8$, the normalisation of the concentration-mass relation for dark matter haloes $f_{\mathrm{h}}$, the normalisation of stellar-to-halo mass relation $M_0$, the characteristic scale of the stellar-to-halo mass relation $M_1$, the slope parameters of the stellar-to-halo mass relation $\gamma_1$ and $\gamma_2$, the scatter between stellar mass and halo mass $\sigma_{\mathrm{c}}$, the normalisation of the concentration-mass relation for distribution of satellite galaxies $f_{\mathrm{s}}$, the power law behaviour of satellites $\alpha_{\mathrm{s}}$, the normalisation constants of the Schechter function $b_0$ and $b_1$, and the Poisson parameter $\mathcal{P}$. Parameter $M_{\mathrm{piv}}$ is the location of the power law used to describe the mass dependence of the Poisson parameter $\mathcal{P}$. Parameters are deemed unconstrained when the marginal probability at $2\sigma$ level exceeds 13\% of the peak probability \citep[see appendix A of][]{Asgari2020}. In cases where one side is constrained we report the $1\sigma$ lower/upper limit. The MMAX estimate is the marginal maximum statistic, reporting the point of maximum marginal posterior distribution to the iso-posterior levels above and below the maximal point. The MAP+PJ-HPD (maximum posterior with projected joint highest posterior density) estimates are calculated following \citet{Joachimi2020}.}
\end{table*}

\subsection{Modelling intrinsic galaxy alignments and magnification}
\label{sec:ia_mag_test}

\begin{figure}
	\centering
	\includegraphics[width=\columnwidth]{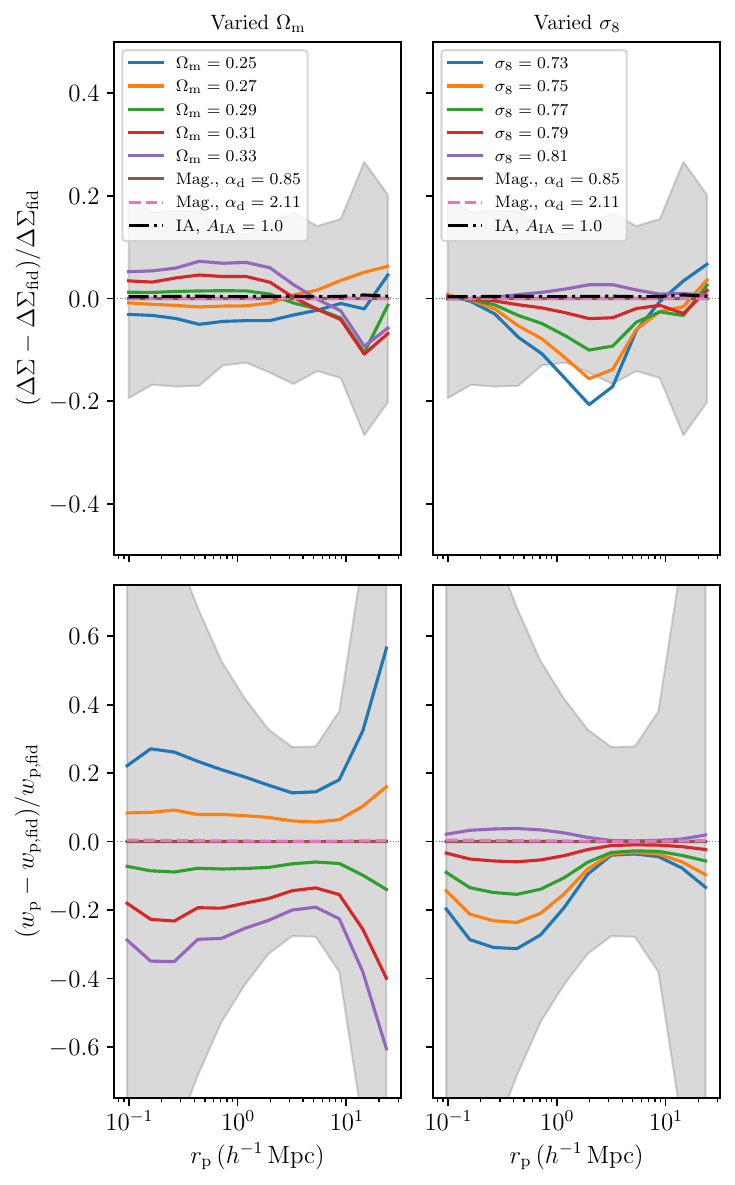}
 	\caption{Relative difference between the lensing signal with and without the contribution of intrinsic alignments and lens magnification. The intrinsic alignment contribution is modelled with $A_{\mathrm{IA}} = 1$, which is a reasonable `worst case' scenario for the KiDS-Bright sample, and the magnification is modelled with $\alpha_{\mathrm{d}} = 0.85$ and $\alpha_{\mathrm{d}} = 2.11$, typical for GAMA-like lenses at a redshift of 0.21 and 0.36. Grey areas show the error on the data. We also show the changes to the galaxy-galaxy lensing and galaxy clustering signals if we change the two main cosmological parameters. The behaviour is only shown for the largest stellar mass bin, as the contribution to magnification and intrinsic alignments is of the same order for all of them.}
	\label{fig:esd_mag}
\end{figure}

We quantify the contribution of lens galaxy magnification and intrinsic galaxy alignments to our data, which we do not account for in our fiducial model. We estimate the contribution of intrinsic alignments (IA) using the NLA model \citep{Bridle2007} by
\begin{equation}
P_{\mathrm{gI}}(k,z) = - A_{\mathrm{IA}} C_{1} \rho_{\mathrm{crit}} \frac{\Omega_{\mathrm{m}}}{D(z)} P_{\mathrm{gm}}(k,z)\,,
\end{equation}
where $A_{\mathrm{IA}}$ is the amplitude of the intrinsic alignment signal, $C_{1}$ is a normalisation constant, $D(z)$ the linear growth factor. We set $C_{1} \rho_{\mathrm{crit}} = 0.0134$, motivated by \citet{Brown2002}. In order to estimate the contribution of intrinsic alignments to the galaxy-galaxy lensing signal, we set $A_{\mathrm{IA}} = 1$, which is a good `worst-case' scenario for our complete magnitude-limited KiDS-Bright sample.  We project the $P_{\mathrm{gI}}$ power spectrum to the galaxy-galaxy lensing signal. We subtract the additional contribution from the intrinsic alignments from the measured lensing signal and compare the relative difference between the corrected and uncorrected data to quantify the impact of neglecting intrinsic galaxy alignments in our theoretical model. Such model is only describing the IA well on large scales and becomes increasingly ad hoc on small scales, but it nevertheless provides for a sensible amplitude estimation. Future studies will use the halo model based approach as presented in \citet{Fortuna2020} and \citet{Georgiou2019}, which links the IA signal to the properties of the galaxies an their parent dark matter haloes. 

The magnification contribution to the total lensing and clustering signal is modelled as 
\begin{equation}
\Delta\Sigma_{\mathrm{mag}}(r_{\mathrm{p}}) = 2 (\alpha_{\mathrm{d}} - 1)\, \Delta\Sigma_{\mathrm{mm}}(r_{\mathrm{p}})
\end{equation}
and
\begin{equation}
w_{\mathrm{p, mag}}(r_{\mathrm{p}}) = 4 (\alpha_{\mathrm{d}} - 1)\, w_{\mathrm{gm}}(r_{\mathrm{p}}) + 4 (\alpha_{\mathrm{d}} - 1)^{2}\, w_{\mathrm{mm}}(r_{\mathrm{p}}) \, ,
\end{equation}
where $\Delta\Sigma_{\mathrm{mm}}(r_{\mathrm{p}})$ and $w_{\mathrm{mm}}(r_{\mathrm{p}})$ are the lensing signal and projected clustering contributions from the convergence power spectrum, respectively \citep{Joachimi2010, Simon2017, Unruh2020}. The $w_{\mathrm{gm}}(r_{\mathrm{p}})$ is the same quantity as the $\Sigma(r_{\mathrm{p}})$, but without the mean density $\overline{\rho}_{\text{m}}$ (see Eq. \ref{Sigma_approx}). We estimate the contribution from lens magnification using the values for a KiDS and GAMA like survey, given by \citet{Unruh2020}. We adopt $\alpha_{\mathrm{d}} = 0.85$ corresponding to a lens redshift of 0.21, and $\alpha_{\mathrm{d}} = 2.11$ corresponding to a lens redshift of 0.36. These values are representative of the magnification effect we would expect for the two highest stellar mass bins in our analysis, with other bins predicted to have a somewhat smaller contribution, due to their redshifts \citep{Unruh2020}. We subtract the magnification contribution from our data to quantify the impact of neglecting magnification in our theoretical model. 

The relative contributions of intrinsic alignments and magnification to the ESD and $w_{\mathrm{p}}$ observables are presented in Fig. \ref{fig:esd_mag}. We present the contribution for the largest stellar mass bin only, as the contribution to magnification and intrinsic alignments is of the same order for all of them. On the same figure we also show the changes to the galaxy-galaxy lensing and galaxy clustering signals if we change the two main cosmological parameters. The contributions of both intrinsic alignments and magnification are well below 1\%, as also found in \citet{Unruh2020} and are well within the error budget of the data. Moreover, they are also subdominant to the changes due to the cosmological parameters of interest. Such contributions have negligible effects on the overall signal, and are unlikely to be a significant source of bias. To some extent the effects cancel each other out, as for the redshifts of our lens galaxies the magnification effect dilutes the signal, while the intrinsic alignments add a similar contribution.

\subsection{Changing the $n_{\rm s}$ prior}
\label{sec:ns_model}

Inspecting Fig.~\ref{fig:full_corner} we note that the marginalised posteriors of our prior-informed cosmological parameter set, $n_{\rm s}$, $\Omega_{\rm b}$ and $h$, have a tendency to push up against one side of the prior.  As the KiDS 2$\times$2pt+SMF data vector is expected to be insensitive to changes in this parameter set, we conclude that this effect arises from projection effects or the MCMC not fully sampling this part of parameter space.  Given that there are no strong degeneracies between this set and the rest of the parameters, and that the set is already well constrained from other studies \citep[see the discussion in appendix B of][]{Heymans2020} we find no motivation to investigate the impact of widening the priors.   Instead we investigated reducing the prior range by adopting a Gaussian prior with the mean and uncertainty fixed to the \citet{PlanckCollaboration2018} constraint for $n_{\rm s}$.  This parameter is the most interesting to investigate of the three, as any tension between small and large scales may be expected to manifest in a biased spectral index constraint \citep{Troester2021}.

The results are presented in Table~\ref{tab:results_ns} and Fig.~\ref{fig:full_corner}.  We find that the marginal contours for the parameters do not change significantly with the addition of a restrictive prior on $n_{\rm s}$.  For example, the MMAX estimate of $S_8$, shifts by $0.13\sigma$, which is consistent with the expected MCMC run-to-run variance \citep{Joachimi2020}.   In future studies we will explore the impact of using more restrictive priors on all externally-constrained parameters which we are insensitive to, noting that this may also help to reduce projection bias for posteriors with many parameters.

\subsection{The effect of unmodelled photometric redshift errors on the projected galaxy clustering measurements}
\label{sec:dilution}

\begin{figure}
	\centering
	\includegraphics[width=\columnwidth]{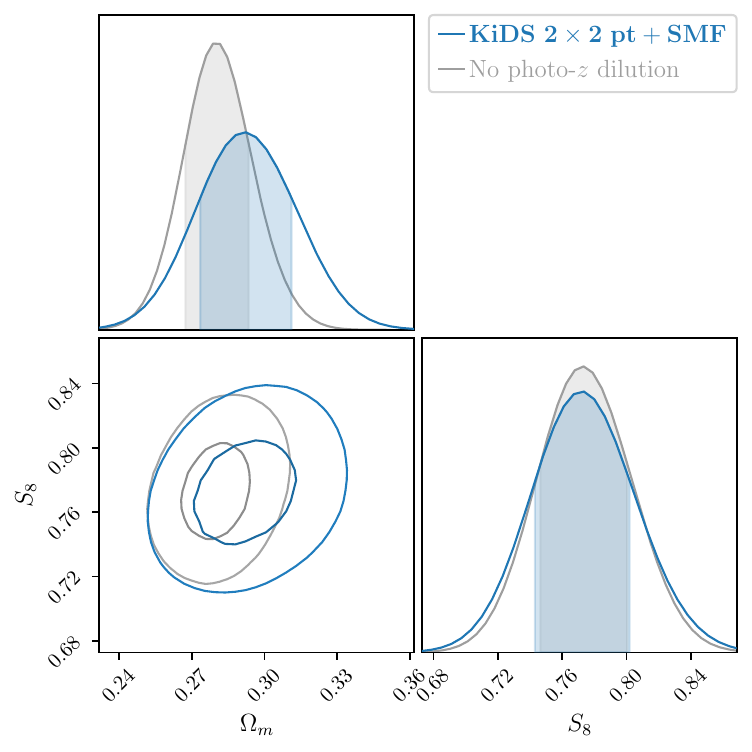}
 	\caption{Marginalised constraints for the joint distributions of $S_8$ and $\Omega_{\mathrm{m}}$, showing 68\% and 95\% credible regions.  We compare our fiducial analysis (blue) with an analysis neglecting the impact of photometric redshift uncertainties which dilute the estimated projected galaxy clustering signal (grey).}
	\label{fig:dilution}
\end{figure}

In Sect.~\ref{sec:proj_clustering} we introduce a free nuisance parameter, $\mathcal{D}$ (Eq.~\ref{eqn:Dfact}), to account for our uncertainty on the amplitude of the true projected galaxy clustering signal.  The expected dilution is a result of unaccounted photometric redshift errors in our theoretical model for $w_{\mathrm{p}}(r_{\mathrm{p}})$.  In Figure \ref{fig:dilution} we compare our fiducial constraints in $S_{8}$ and $\Omega_{\mathrm{m}}$ with the constraints from an analysis where the photometric redshift dilution effect is neglected and $\mathcal{D}=1$. We find that while the omittance of the photometric redshift dilution effect does not impact the $S_{8}$ constraints, $\Omega_{\mathrm{m}}$ becomes biased and more constrained. This is expected as the galaxy clustering is more sensitive to $\Omega_{\mathrm{m}}$ compared to $\sigma_{8}$ (cf. Fig.~\ref{fig:esd_mag}).  This motivates future work to improve the estimator for, or theoretical modelling of, the projected galaxy clustering signal in the presence of photometric redshift errors, following \citet{Joachimi2011, Chisari2014}.

\end{appendix}

\end{document}